\documentclass[a4paper,11pt]{article}
\pdfoutput=1 

\usepackage{jcappub} 

\usepackage[T1]{fontenc} 

\usepackage{hyperref}

\title{\boldmath How to optimally combine pre-reconstruction full shape and post-reconstruction BAO signals}

\author{H\'ector Gil-Mar\'in}

\affiliation{Dept. F\'isica Qu\`antica i Astrof\'isica, Institut de Ci\`encies del Cosmos (ICCUB), Facultat de F\'isica, Universitat de Barcelona (IEEC-UB), Mart\'i i Franqu\`es, 1, E08028 Barcelona, Spain}

\emailAdd{hectorgil@icc.ub.edu}

\abstract{We review the different approaches for combining the cosmological information from the full shape of the pre-reconstructed power spectrum -  usually referred as redshift-space distortion (RSD) analysis - and from the baryon acoustic oscillation (BAO) peak position in the post-reconstructed power spectrum with the aim of finding the optimal procedure. We focus on combining the pre- and post-reconstructed derived quantities at different compression levels: 1) the two-point summary statistics, the power spectrum multipoles, $P^{(\ell)}(k)$; 2) the compressed BAO variables, $\alpha_{\parallel,\perp}$; and 3) an hybrid approach between 1) and 2). We apply these methods to the publicly available eBOSS Luminous Red Galaxy catalogues, for both data and synthetic EZ-mocks. We find that the three approaches result in very consistent posteriors when the appropriate covariance matrix estimator is used. 
On average, the combination at $P^{(\ell)}(k)$ level retrieves $5-10\%$ tighter constraints than the other two approaches, demonstrating that the standard approach of combining at the level of the BAO variables is nearly optimal.
We conclude that combining both BAO post-reconstructed and full shape pre-reconstructed signals for the one single data realization at the level of the summary statistics is faster, as it does not require
running the whole pipeline on the individual mocks, and brings a moderate $10\%$ improvement, with respect to the other two studied methods. Moreover, we check for potential systematics, such as, the way the matrix is built and the effect of the finite number of mocks on the likelihood estimator and find none of these have a significant impact in the final results. Combining the pre- and post-reconstruction signals at the level of the summary statistics is an attractive, faster and accurate method to be used in future and on-going spectroscopic surveys.
}

\begin{document}
\maketitle
\flushbottom

\section{Motivation}
\label{sec:intro}

The Baryon Acoustic Oscillations (BAO) signal in the galaxy distribution is one of the most robust probes in late-time cosmology. The BAO has been, along with the redshift space distortions (RSD) signature, the cornerstone for performing cosmology inference with massive galaxy spectroscopic surveys for the last decade. The adoption of the BAO as standard ruler has allowed the calibration of the redshift of distant galaxies and to determine their true comoving distances. The application of these technique to the largest 3D spectroscopic galaxy maps up-to-date elaborated by the Sloan Digital Sky Survey (SDDS) programs - Baryon Oscillator Spectroscopic Survey (BOSS) and extended BOSS (eBOSS) - has conducted to the measurement of the expansion history and growth of perturbations over 11,000 million year of cosmic evolution with unprecedented precision and accuracy \cite{alam_clustering_2017,eboss_collaboration_dr16}. In addition, the RSD signal has provided constraints on General Relativity as the theory of gravity at inter-galactic scales.

The state-of-the-art spectroscopic BAO-only analyses are already able to detect the effect of Dark Energy at $8\sigma$ significance, confirming the paradigm of the acceleration of the expansion of the Universe established by pioneering observations of type Ia supernovae \cite{riess,perlmutter}, and supporting the $\Lambda$CDM model as the standard model in cosmology. Recently, also the broadband {\it Shape} of the power spectrum \cite{ShapeFit} has emerged as a complementary observable to the standard BAO and RSD signals, being able to add extra information on primordial non-Gaussianities and on the sum of the neutrino masses \cite{BriedenPRL21}, which allows for internal stress-tests on the $\Lambda$CDM model. 

Although these three distinct features, BAO, RSD and {\it Shape}, can simultaneously be measured from the resulting summary statistics of the galaxy catalogues (usually just the two-point statistics, or the power spectrum in Fourier space), in modern analyses it is customary to apply the reconstruction technique \citep{Eis2007} and generate an additional data set, the post-recon catalogue, and consequently additional post-recon power spectra. The reconstruction technique can undo the effect of bulk flows caused by peculiar velocities, sharpening the significance of the BAO peak in the reconstructed two-point statistics. This process can be seen as a `Gaussianization' of the original galaxy catalogue. It is precisely because of this that the resulting two-point statistics from these two catalogues, pre- and post-recon, contain significantly different information. The post-recon power spectrum holds information which not only comes from the two-point statistics from the original pre-recon catalogue, but also from its three- and four-point statistics \cite{Wangetal21}. In this fashion, the reconstruction process moves information from higher-order moments back to the two-point order statistics. Hence, any analysis which aims to use both pre- and post-recon two-point information needs to account for these two correlated, but distinct, catalogues.

In order to provide a single set of cosmological parameters derived from the observations we need to `re-combine' the information coming from these two  catalogues. From the pre-recon catalogue a compressed set of variables can be measured: two variables containing BAO information along and across the line-of-sight, $\alpha_\parallel$ and $\alpha_\perp$, respectively;\footnote{Here $\alpha_{\parallel,\perp}$ accounts for the longitudinal and transverse dilation scales and the ratio of sound horizon scales between the true cosmology and the template cosmology. See for e.g., section 3 of \cite{ShapeFit} for a full description of these and other cosmological parameters.} the logarithmic growth of structure, $f$; and the amplitude of matter fluctuations at scales of $8\,{\rm Mpc}h^{-1}$, $\sigma_8$ (usually constrained under the combination of $f\cdot\sigma_8\equiv f\sigma_8$); and broadband shape information, $m$ (see for e.g., \cite{ShapeFit,ShapeFitPT} for a description of this parameter and its connection to the transfer function). On the other hand, from the post-recon catalogues only the BAO peak information is extracted, and therefore we can only infer $\alpha_\parallel$ and $\alpha_\perp$. In principle, the post-recon power spectrum could also be used to extract information on $f\sigma_8$ and on $m$, although an accurate model for the full shape of the post-recon field would be needed. In any case, both catalogues need to be used and combined correctly and consistently for an accurate and precise extraction of cosmological information. 

In this paper we study the impact of different types of approaches used in the literature when combining pre- and post-recon power spectrum signals. We denote this combination through the symbol $\wedge$. For simplicity, and as it is the case of most state-of-the-art analyses, we will employ the pre-recon catalogues to determine the full shape compressed variables, $D^{\rm FS}=\{\alpha_\parallel,\alpha_\perp,f\sigma_8,m\}^{\rm pre}$, and the post-recon catalogues to determine the BAO compressed variables, $D^{\rm BAO}=\{\alpha_\parallel,\alpha_\perp\}^{\rm post}$. Also, for simplicity (and the limitation on the available number of synthetic mock catalogues used for generating the covariance) we will solely focus on the power spectrum multipoles, not using any higher-order statistics. These mock catalogues consist of a large number of simulations which are run using fast techniques and that are able to emulate the features of the actual data catalogue, both in survey geometry and clustering properties.

Most of the BOSS and eBOSS collaboration works employ the approach of performing independent BAO post-recon and full shape pre-recon analyses. Both compressed data-vectors, $D^{\rm BAO}$ and $D^{\rm FS}$, are then combined {\it a posteriori} into a single set of cosmological parameters using the parameter-covariance extracted from the mocks, $D^{\rm BAO}\wedge D^{\rm FS}\rightarrow D^{\rm FS+BAO}$. This approach has the drawback that the whole BAO and full shape pipelines need to be run a large number of times on the pairs `pre-recon post-recon' mocks in order to derive this parameter-covariance matrix. An alternative approach is used in some eBOSS papers \cite{Gil-Marin:2020bct,demattiaetaleboss21} where both post-recon BAO and pre-recon RSD analyses are simultaneously performed. In this approach the full pre- and post-recon power spectrum data vector is simultaneously employed to obtain a single compressed set of variables, $P_{\rm pre}^{(\ell)}(k)\wedge P_{\rm post}^{(\ell')}(k')\rightarrow D^{\rm FS+BAO}$. The advantage is that the analysis pipeline only needs to be run one time, on the data catalogue, as the covariance exclusively comes from the pre- and post-recon $P^{(\ell)}(k)$ measurements. Also, for small samples, the BAO detection in the pre-recon catalogue may be poor, resulting in an almost unconstrained $\alpha_{\parallel,\,\perp}$, but by simultaneously fitting pre- and post-recon catalogue the detection of BAO (and the constraints on $\alpha_{\parallel,\,\perp}$) improves \cite{demattiaetaleboss21}. 

 An hybrid approach is to combine the post-recon compressed BAO variables with the pre-recon spectra into a single data-vector, $D^{\rm BAO}\wedge P_{\rm pre}^{(\ell)}(k) \rightarrow  D^{\rm FS+BAO}$. This is the approach followed by \cite{2020JCAP...05..032P,2021arXiv211204515P} using BOSS data. Recently, \citep{shi-fan} also presented an analysis on BOSS data using as data-vector the pre-recon power spectrum, $P_{\rm pre}^{(\ell)}(k)$, and post-recon correlation function, $\xi_{\rm post}^{(\ell)}(s)$. In this approach the BAO feature is naturally isolated in configuration space and therefore the pre-post covariance is better behaved under inversion.

In this paper we test different approaches to derive a single set of compressed cosmological parameters from the two-point summary statistics of the pre- and post-recon catalogues, and compare the results, both in terms of signal and errors. As a benchmark we use the publicly available SDSS-IV Luminous Red Galaxy (LRG) sample from the BOSS+eBOSS catalogues, in the redshift range $0.6\leq z \leq 1.0$, and with an effective redshift of $z_{\rm eff}=0.70$. Details on how the data catalogues have been constructed can be found in \cite{ebossLRG_catalogue}.  The reconstruction algorithm to generate the post-reconstruction analyses is based on the works by \cite{burden_efficient_2014,burden_reconstruction_2015}. We employ 1000 realizations of the EZ mocks \citep{Ezmocks} with the same geometry and redshift selection window as the data, for constructing reliable covariances which account for the most important observational effects.\footnote{The data catalogues used in this paper are publicly available \href{https://data.sdss.org/sas/dr16/eboss/lss/catalogs/DR16/}{here}. The measured pre- and post-recon power spectra are available \href{https://svn.sdss.org/public/data/eboss/DR16cosmo/tags/v1_0_1/dataveccov/lrg_elg_qso/LRG_Pk/}{here}. An implementation of the reconstruction algorithm can be found \href{https://github.com/julianbautista/eboss_clustering}{here}.}

This paper is structured as follows. In \S\ref{sec:methodology} we describe the several approaches studied for combining pre- and post-reconstructed information. In \S\ref{sec:results} we present the results of applying these approaches on mocks and data. In \S\ref{sec:sys} we test for different potential systematic effects from the model and covariance. Finally in \S\ref{sec:conclusions} we present the conclusions.

\section{Methodology}\label{sec:methodology}
In this work we focus on Fourier space products. We consider the power spectrum of the pre- and post-recon fields, $P_{\rm pre}(k)$ and $P_{\rm post}(k)$, respectively, measured from eBOSS catalogues, as described in \citep{Gil-Marin:2020bct}. In the case of a Gaussian field, the power spectrum represents a lossless compression of the information-content of the field. In practice, the actual galaxy field is non-linear and presents a strong non-Gaussian component in its signal due to non-linear evolution, and therefore, relevant cosmological information is contained in the higher-order statistics. By applying the reconstruction technique the galaxy field becomes more Gaussian, and part of the signal in the higher-order moments moves back into the power spectrum. Thus, by using both pre- and post-recon power spectra it is possible to increase the amount of information with respect to the pre-recon power spectra alone.

\subsection{Power spectrum models}

We use the power spectrum multipoles (monopole, quadrupole and hexadecapole) of the pre-recon field to perform a full shape analysis. This type of analysis requires modeling the whole shape of the power spectrum in a certain range of scales ($0.02\leq k\,[{\rm Mpc}^{-1}h]\leq 0.15$ for this paper). In order to do so, we apply the Resummed Eulerian Perturbation Theory prediction at 2-loops (2L-RPT, \cite{Gil_Mar_n_2012}), where the redshift space distortions are implemented as described by the Taruya-Nishimichi-Saito model (TNS, \cite{Taruya:2010mx}). We consider four cosmological parameters\footnote{Along this paper we refer as `cosmological parameters' the physical parameters inferred by the analyses that enclose cosmology information, which in turn, can be converted in the traditional cosmological parameters given a cosmological model.} for this model, which we refer to as the full shape set of compressed variables, $D^{\rm FS}=\{\alpha_\parallel,\alpha_\perp,f\sigma_8,m \}$.
We also marginalize over four extra nuisance parameters per galactic cap (northern and southern), such as, the linear and non-linear biases, the shot noise amplitude and the Fingers-of-God damping parameter. Further details on this model can be found in \citep{Gil-Marin:2020bct}. 

In addition, we use the power spectrum multipoles (monopole and quadrupole) of the post-recon field and perform a BAO analysis within the $k$-range, $0.02\leq k\,[{\rm Mpc}^{-1}h]\leq 0.30$. We use the model presented in Eqs. 22-23 of \citep{Gil-Marin:2020bct} (also see \cite{Beutler:2016ixs}), consisting of a broadband and an oscillatory terms. The broadband term has $N+1$ free parameters per galactic cap, and per multipole, $\{B,\, A_1,\,A_2\,\ldots\,A_N\}$. Additionally an extra effective nuisance parameter, $\beta$, common for all patches and multipoles, is added. The oscillatory term has two BAO-scaling parameters, which constitute the compressed set of cosmological parameters, $D^{\rm BAO}=\{\alpha_\parallel\,\alpha_\perp\}$. Moreover, two BAO-damping parameters are also added to the model, although we keep them fixed to their best-fit value of the mocks. Exhaustive analyses on the impact of this assumption were already presented in \citep{Gil-Marin:2020bct}. 

Thus, in terms of relevant information for cosmology, the band-power information of the pre- and post-reconstructed summary statistics is squeezed into the following compressed data-vectors,
\begin{eqnarray}
   \label{eq:compressionFS} P_{\rm pre}^{(\ell)}(k_i) &\longrightarrow& D^{\rm FS}= \{\alpha_\parallel,\alpha_\perp,f\sigma_8,m\}, \\
  \label{eq:compressionBAO}  P_{\rm post}^{(\ell)}(k_i) &\longrightarrow&D^{\rm BAO}=\{\alpha_\parallel\,\alpha_\perp\}, 
\end{eqnarray}
where for the pre-recon set of power spectrum multipoles this represents a compression from $2\times39$ band-powers (among all $k$-vectors, multipoles) down to just 4 relevant quantities; and for the post-recon power spectrum from $2\times56$ down to just 2.

The set of relevant parameters, $\{\alpha_\parallel,\alpha_\perp,f\sigma_8,m\}$ should be converted into physical parameters to be interpreted later within a cosmological model. The geometric parameters, $\alpha_\parallel$ and $\alpha_\perp$ distort the true scales with respect to the observed ones given the fixed cosmology (also referred as fiducial cosmology) used to transform redshifts into distances. In this case, $\Omega_m^{\rm fid}=0.31$. A difference in values between the true cosmology and the fiducial cosmology produces both an isotropic dilation of scales and an Alcock-Paczynski effect. In addition the full shape and BAO analyses of this work are performed at a fixed template, with a fixed sound horizon scale, given by the fiducial cosmology of the template,\footnote{Although not strictly necessary, we choose the fiducial cosmology of the template to be the same as the fiducial cosmology to convert redshifts into distances.} $r_s^{\rm fid}$. The different values of sound horizon scales, between the template cosmology and the true value, make the BAO scale to shift isotropically, in identical manner as the geometric isotropic dilation. Taking into account these considerations, the dilation parameters are related to the Hubble distance parameter, $D_H$ and the comoving angular diameter distance, $D_M$, as \cite{1979Natur.281..358A}, 
\begin{equation}
\frac{D_H(z)}{r_s}\equiv\frac{c}{r_sH(z)}=\frac{c}{r_s^{\rm fid} H^{\rm fid}(z)} \alpha_\parallel(z)=\left[\frac{D_H(z)}{r_s}\right]^{\rm fid}\alpha_\parallel(z)
\end{equation}

\begin{equation}
    \frac{D_M(z)}{r_s}\equiv \int_0^z\frac{c\,dz'}{r_sH(z')}=\int_0^z\frac{c\,dz'}{[r_sH(z')]^{\rm fid}} \alpha_\perp(z)=\left[ \frac{D_M(z)}{r_s} \right]^{\rm fid}\alpha_\perp(z)
\end{equation}
where $c$ is the speed of light and $H$ the Hubble parameter. 

The measurement of the growth of structure parameter, $f$, is performed at a fixed template without varying the amplitude of linear power spectrum terms, this is at fixed $\sigma_8$ value. Therefore this measured growth of structure, $\widetilde{f}$, is intrinsically biased because the value of the amplitude of fluctuations is, in general, different between the fiducial cosmology and the one studied. Since $f$ and $\sigma_8$ are very degenerate, this effect is fixed by simply taking $f\sigma_8(z)= \widetilde{f}(z)\cdot\sigma_8^{\rm fid}(z)$.
There is, however, an additional correction we should consider. In the above definition, $\sigma_8$ corresponds to the fluctuation of the matter field smoothed in spheres of $8\,{\rm Mpc}h^{-1}$, according to `observed' units, different in general from the underlying true units. One needs to include the effect of the isotropic dilation scales in the definition of the size of the spheres from the fiducial cosmology in order to account for this effect  \cite{Gil-Marin:2020bct},
\begin{equation}
    f\sigma_8(z)=\widetilde{f}(z)\cdot\sigma^{\rm fid}_8(z;\alpha_{\rm iso}),
\end{equation}
where $\alpha_{\rm iso}\equiv[\alpha_\parallel\alpha_\perp^2]^{1/3}$, and $\sigma^{\rm fid}_8(z;\alpha_{\rm iso})$ is given by Eq. 40 of \cite{Gil-Marin:2020bct}.

The measured shape parameter, $m$, contains information from the early-time processes in the universe through the slope of the transfer function in scales between the BAO scales and the equality scale. In general, the relation with the transfer function is given by Eq. 3.14 of \cite{ShapeFit}.

\subsection{Combining pre- and post-recon information}

So far we have described the full shape and BAO analyses independently. However, we wish to combine them into a single compressed variables vector, $D^{\rm FS+BAO}=\{\alpha_\parallel,\alpha_\perp,\widetilde{f},m\}$. As we have already anticipated, the pre- and post-recon power spectra are highly correlated and in order to compute $D^{\rm FS+BAO}$ we need to account for that correlation. We do so by using the EZ mocks, consisting of 1000 independent realizations per galactic cap. In practice the two galactic caps are considered independent, and their likelihoods are multiplied as, $\mathcal{L}_{{\rm n}+\rm{s}}=\mathcal{L}_{\rm n}\cdot\mathcal{L}_{\rm s}$. Thus, the covariance matrix of the whole north+south region is estimated from 1000 distinct realizations (see \S\ref{sec:cov_estimation} for details) which provide the correlation among $k$-bins and $\ell$-multipoles of a given summary statistic, $P_{\rm pre}^{(\ell)}(k_i)$ or $P_{\rm post}^{(\ell)}(k_i)$, needed to perform the fit and the compression described in the Eqs. \ref{eq:compressionFS}-\ref{eq:compressionBAO}. In addition, the mocks are also used to infer the correlation among pre- and post-recon elements (i.e., compressed variables or summary statistics). 

\begin{figure}[ht]
    \centering
    \includegraphics[scale=0.205]{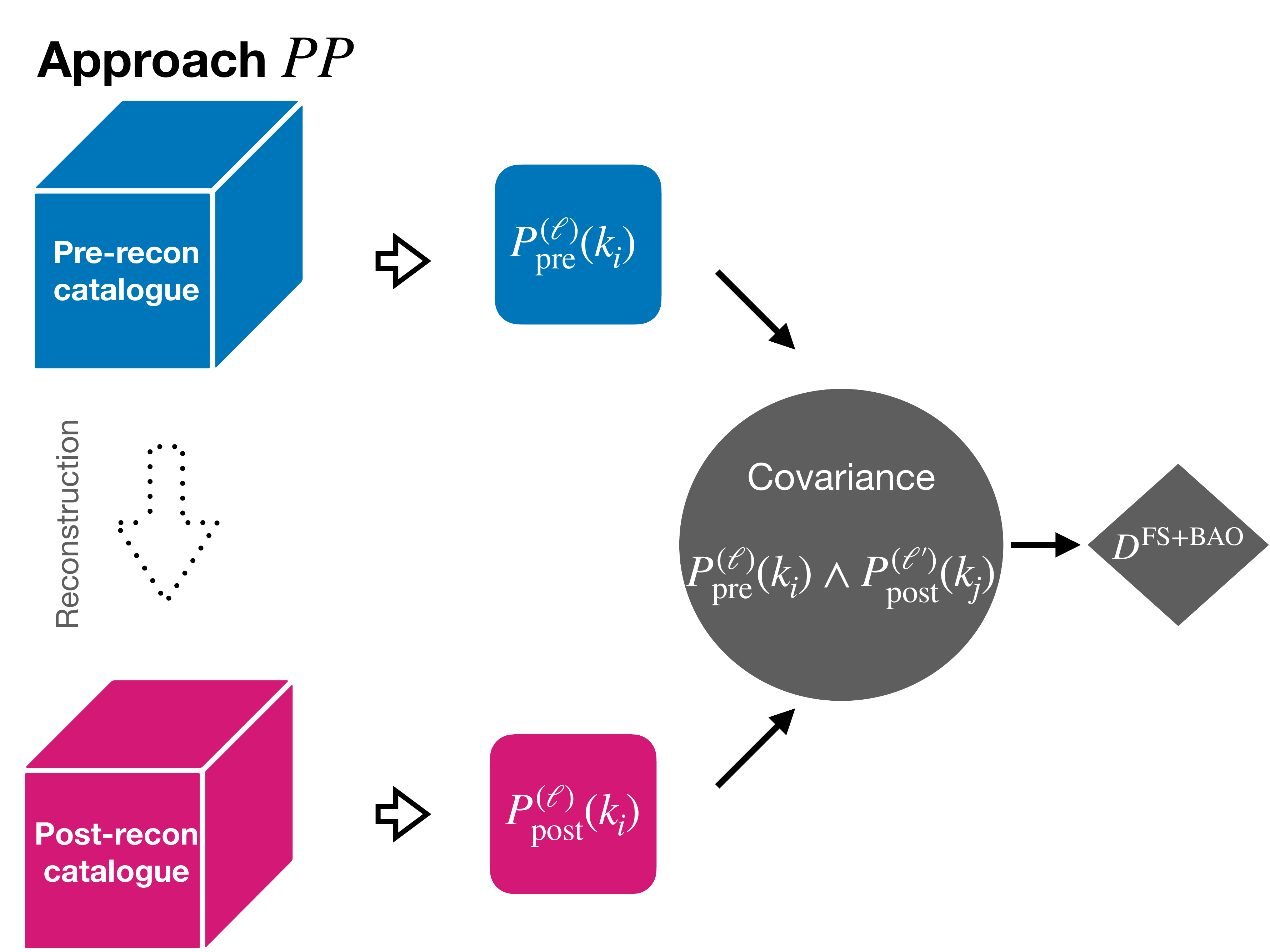}
        \includegraphics[scale=0.255]{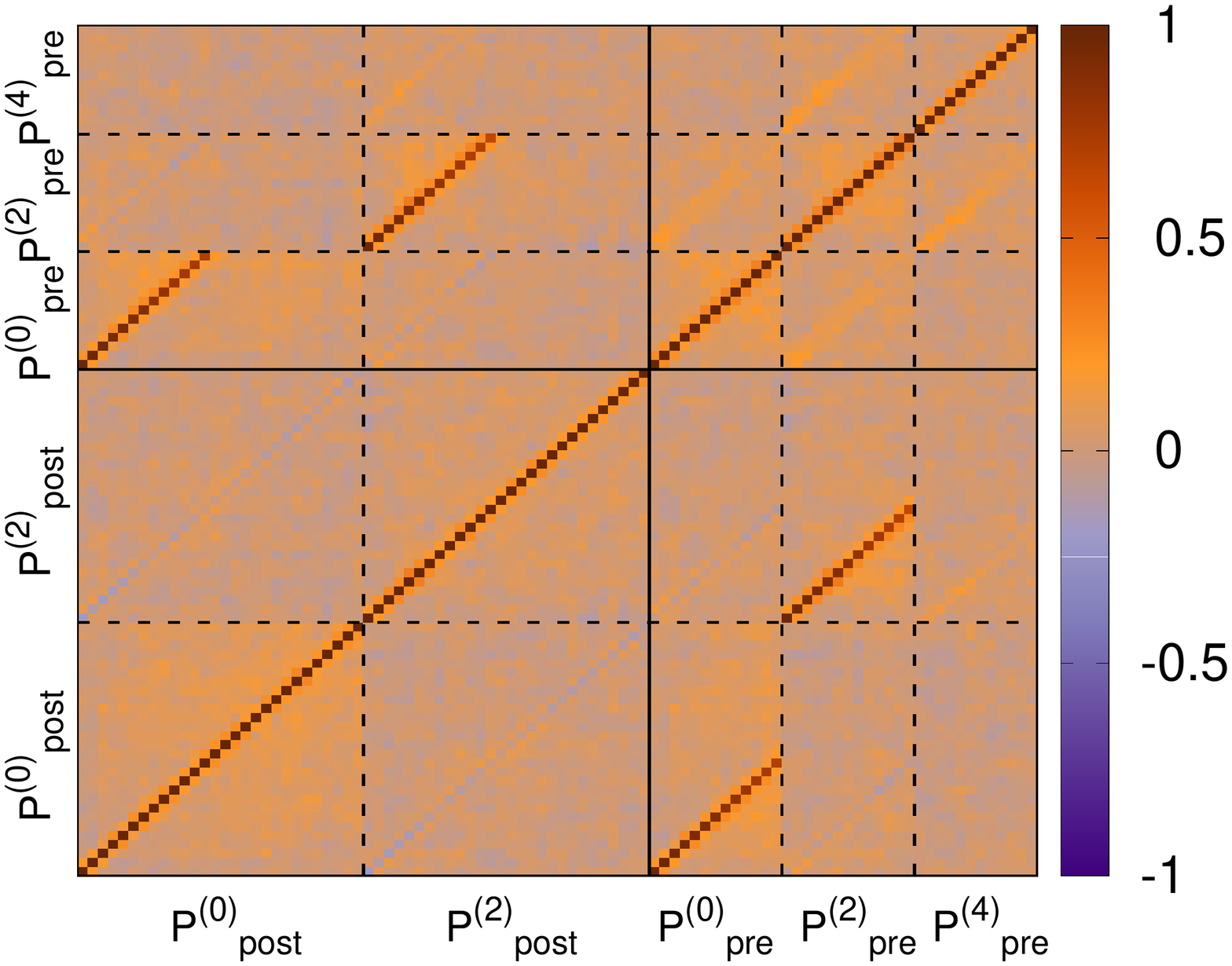}
   
       \includegraphics[scale=0.205]{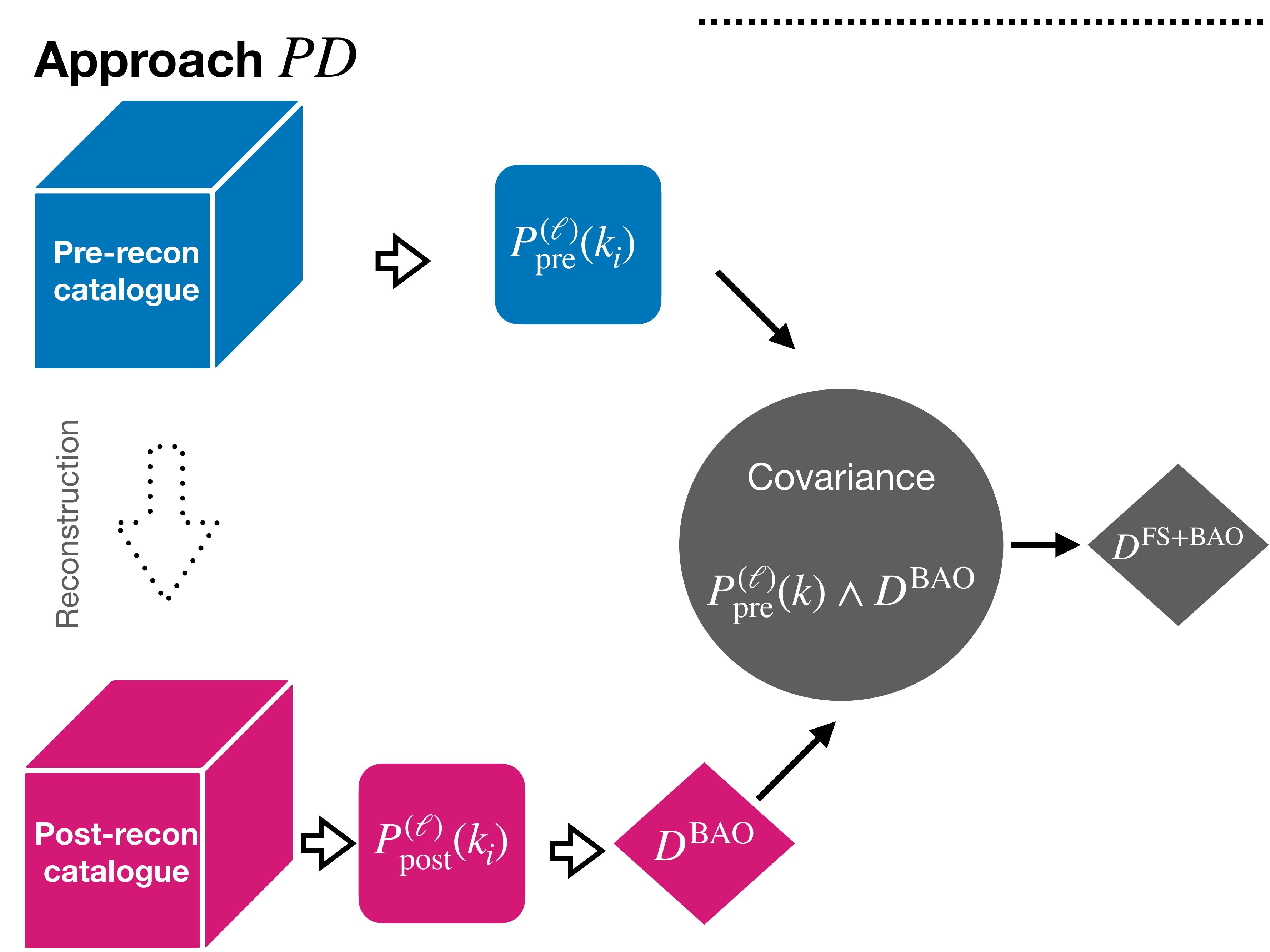}
        \includegraphics[scale=0.255]{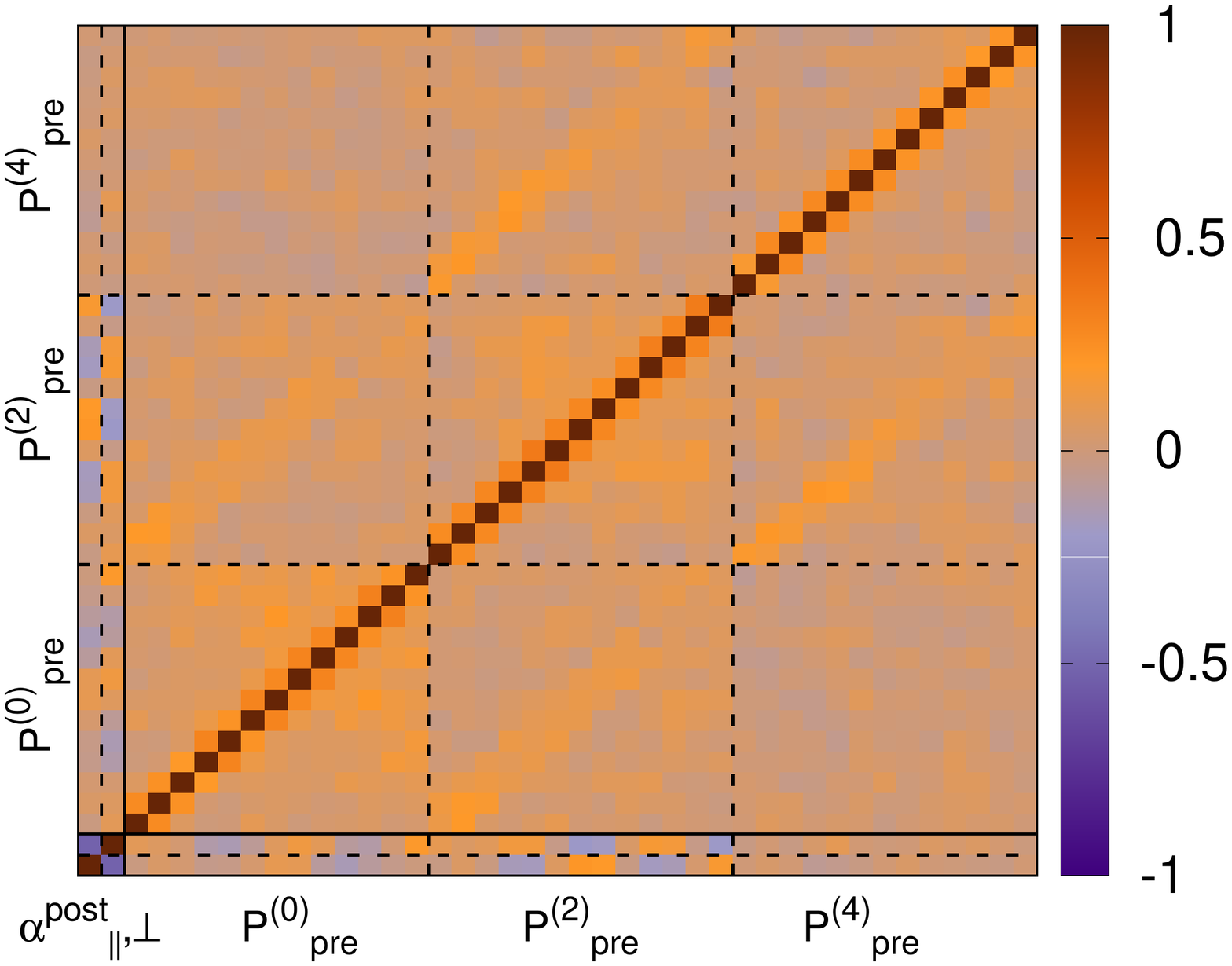}

       \includegraphics[scale=0.205]{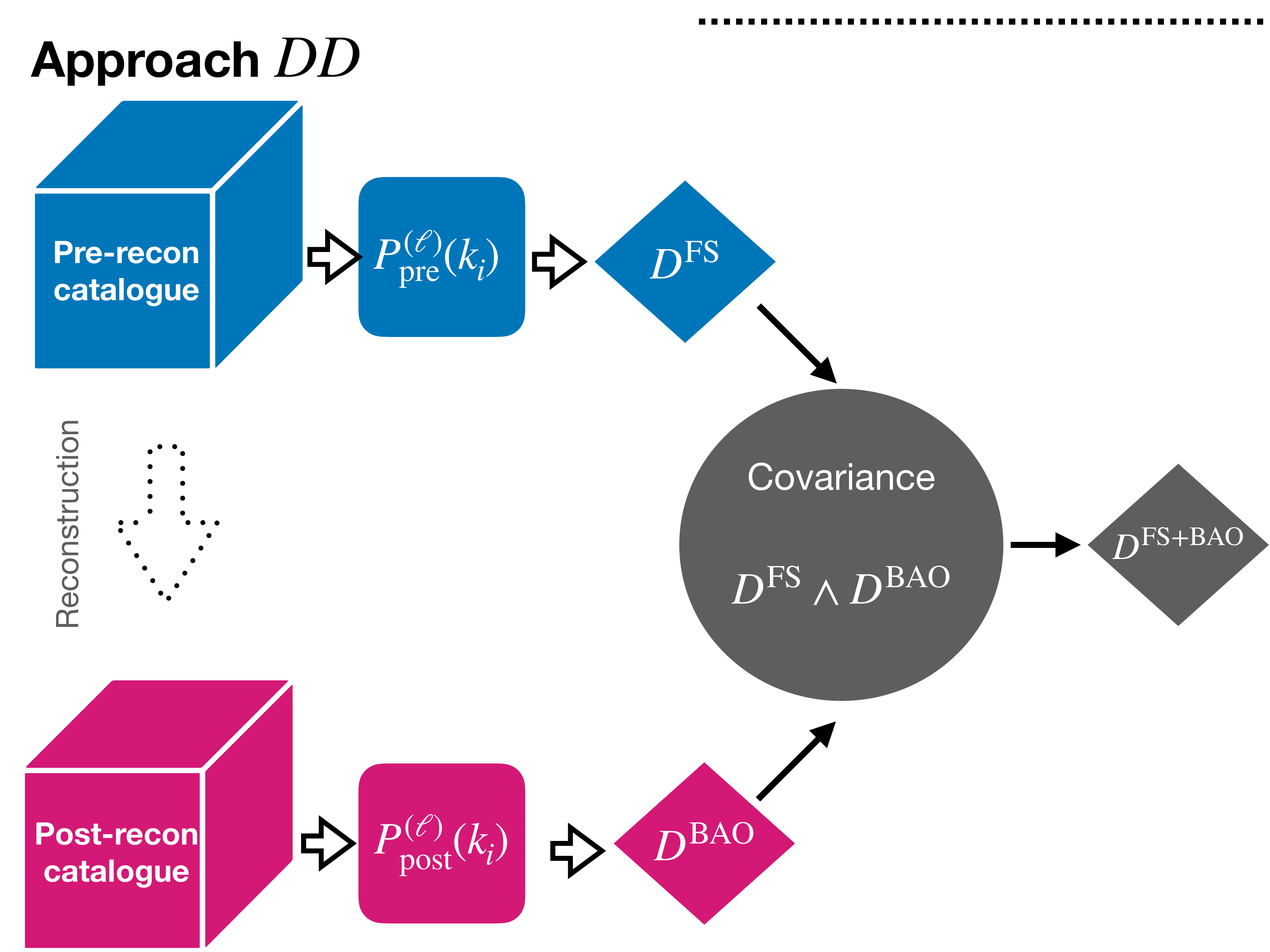}
        \includegraphics[scale=0.255]{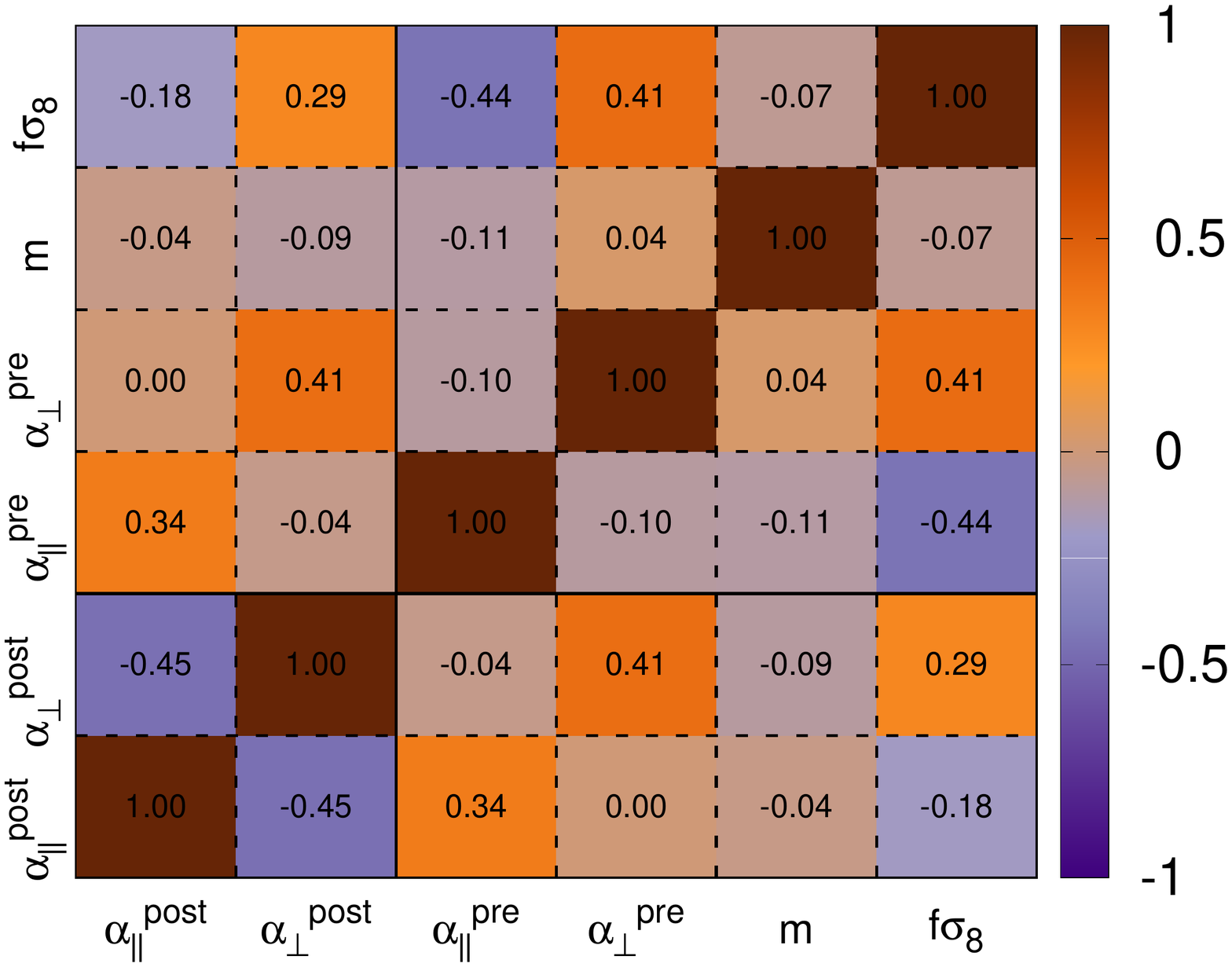}

    \caption{Overview of the approaches considered in this paper when combining pre- and post-recon catalogues. On the left, each row displays the scheme of how the pre- (blue) and post-recon (magenta) catalogues are combined into a single set of compressed parameters, $D^{\rm FS+BAO}$ (in gray). On the right the corresponding covariance used in the gray circle of each corresponding row (only the cross-correlation coefficients for the northern galactic cap are displayed). In the first row, the $PP$-approach uses the covariance at the summary statistics level to perform a simultaneous fit of the compressed parameters. In the second row the $PD$ approach performs a compression of the post-recon statistics, $D^{\rm BAO}$ to later combine it with the pre-recon power spectrum and derive the combined compressed set of variables. In the third row the $DD$ approach performs two independent data compression, $D^{\rm FS}$, $D^{\rm BAO}$, and combines them in order to get a single set of compressed variables. For the $PP$, $PD$ and $DD$ approaches the full data-vectors consist of 95, 41 and 6 elements, respectively}
    \label{fig:esquema}
\end{figure}

We consider the combination of pre- and post-recon information at different levels: the compressed statistics, $DD$;  the summary statistics, $PP$; and a mix of them, $PD$. Below and in Fig.~\ref{fig:esquema} these three approaches are described. 

\begin{itemize}
    \item Summary statistics, $P_{\rm pre}\wedge P_{\rm post}$ (approach $PP)$. We consider the pre- and post-recon summary statistics, $P^{(\ell)}_{\rm pre}(k)$ and $P^{(\ell)}_{\rm post}(k)$ as a single data-vector. Using the power spectra measured from the mocks we build their cross covariance and simultaneously fit the whole data-vector with the appropriate pre- and post-reconstruction models, to obtain a consistent set of $\{\alpha_\parallel,\alpha_\perp, \widetilde{f}, m\}$. 
    
    \item Compressed variables, $D^{\rm FS}\wedge D^{\rm BAO}$ (approach $DD$). We treat the two summary statistics, $P^{(\ell)}_{\rm pre}(k)$ and $P^{(\ell)}_{\rm post}(k)$, independently and perform the full shape and BAO type of analyses, respectively, and obtain two sets of compressed variables, $D^{\rm FS}$ and $D^{\rm BAO}$. 
    We run this pipeline for each $i$-realization of the 1000 mocks, and we use these 1000 pairs of $D_i^{\rm FS},\,D_i^{\rm BAO}$ to build up the cross-covariance matrix between $D^{\rm FS}$ and $D^{\rm BAO}$. With this matrix we are able to reduce the dimensionality of this combined data-vector from 6 to 4. This can be done by linearly combining the redundant parameters (the $\alpha$'s in this case) and determining the linear weights by minimising the variance of the resulting consensus variable (Lagrange multipliers technique), or combining the posteriors assuming that the resulting likelihood among individual measurements is the same as the one from the original data-vector \citep{sanchez_comb}. Later in this section we will revisit the question on how to build this full covariance matrix from the mocks. 
    
    \item Hybrid data-vector, $P_{\rm pre}\wedge D^{\rm BAO}$ (approach $PD$). We consider the BAO compressed variables, $D^{\rm BAO}$ obtained from the fit to $P^{(\ell)}_{\rm post}$. Using the mocks we build up a matrix for the data-vector formed by $D^{\rm BAO}$ and $P_{\rm pre}^{(\ell)}(k_i)$. We then fit the pre-recon model for $P^{(\ell)}_{\rm pre}$ plus two extra inputs for the full data-vector, the $\alpha_\parallel$ and $\alpha_\perp$ values obtained from the BAO only fit. 
    This approach is very similar to perform a fit to $P^{(\ell)}_{\rm pre}(k)$ using priors on the values of alphas coming from the compressed variables of $D^{\rm BAO}$, but unlike the prior case, this method fully accounts for correlations between the $\alpha$'s and the power spectra bins. 

\end{itemize}

Note that these three approaches are all formally consistent and should all produce identical results in the ideal case of 1) lossless compression and 2) noiseless and Gaussian covariance. In a real case scenario these two conditions might not hold in detail, and therefore, potential differences among them could arise. In the next section we will discuss the results of these approaches when applied to realistic mocks and data. 

Before moving on to the results it is important to discuss some subtleties related to how the covariance needed for the analyses are derived. 
When considering any of the methods described above ($PP$, $PD$ or $DD$)  their full covariance is divided in essentially three distinct blocks: two diagonal pre- and post-recon blocks, and one off-diagonal block describing the pre-post mixing (these three type of blocks are shown divided by solid black lines in the covariances of Fig.~\ref{fig:esquema}).
Thus, schematically, the covariance blocks are,\footnote{Note that the Eq. \ref{eq:matrixC} elements are listed as the matrix elements plotted in Fig.~\ref{fig:esquema} for a better comparison. Consequently the diagonal elements are not placed from the usual top-left to bottom-right, but instead from top-right to bottom-left} \footnote{In this work we will always work under the assumption that the likelihoods follow Gaussian statistics.} 
\begin{equation}
\label{eq:matrixC}
C_{ij}=
\begin{bmatrix}
& & \sigma^{({\rm BAO-FS})}_{ij} &  & \sigma^{({\rm FS})}_{ij} & & \\
& & \sigma^{({\rm BAO})}_{ij} &  & \sigma^{({
\rm BAO-FS})}_{ij} & & \\
\end{bmatrix},
\end{equation}
where each of the terms of the matrix can be decomposed as,
\begin{equation}
\label{eq:matrix}    \sigma^{(m)}_{ij}=\sigma^{(m)}_i\sigma^{(m)}_jr^{(m)}_{ij}.
\end{equation}
The $r_{ij}$ elements are the cross-correlation coefficients, which are unity when $i=j$, $\sigma_{ii}^{(m)}={\sigma_i^{(m)}}^2$. 
Note that for the $PP$ approach all the $\sigma_{ij}$ terms represent the covariance of the power spectrum, whereas for the $DD$ approach they represent the covariance of the compressed variables $D^{\rm BAO}$ and $D^{\rm FS}$. This is a notable difference, which impacts the way the covariance should be inferred. 

The right panels of Fig.~\ref{fig:esquema} display the cross-correlation coefficients, $r_{ij}$ extracted from the mocks, for the northern galactic cap. For the $PP$ approach (first row) we see how the off-diagonal elements of both $P_{\rm pre}$ and $P_{\rm post}$ of the same scale and multipole present a high correlation. For the $PD$ approach (second row) we see how the correlation between $P_{\rm pre}$ and the $\{\alpha_\parallel,\,\alpha_\perp\}^{\rm post}$ present a distinctive pattern for the monopole and quadrupole. Further investigation of this can be found in Appendix \ref{sec:PDterms}. For the $DD$ approach we explicitly see the correlation among the compressed parameters before and after reconstruction. The panel displays a correlation of $+0.34$ and $+0.41$ between the pre- and post-recon $\alpha_\parallel$ and $\alpha_\perp$, respectively, as well as a correlation of $-0.18$ and $+0.29$ between $f\sigma_8$ and the post-recon $\alpha_\parallel$ and $\alpha_\perp$.

When it comes to estimating the terms of the covariance for the $PP$ approach, one can either use mocks or a theoretical modelling (see for e.g., \cite{wadekar2020}). In each of these two cases, the derived covariance will be identical (or `fixed') for all the mock realizations, as both the mock- and the theory-approach intrinsically assume that the variance of $P^{(\ell)}(k_i)$ is independent of the realization. The motivation being that independent samples (with different initial conditions), but statistically equivalent (with the same cosmology and epoch) are described by the same covariance matrix.  This is rooted on the assumption that the variance of the initial conditions overdensity field is independent of the random phases of that field. Independent realizations of a given sample share the same variance of their initial conditions power spectrum, but different random phases in their actual overdensity field. Since the variance is not affected by the randomness of the phases, the variance (and covariance) is the same for all realizations. This is indeed the case for all the simulations used in this paper, and we take it as a reasonable assumption for the initial conditions of the Universe.   

\begin{figure}[ht]
    \centering
    \includegraphics[scale=0.18]{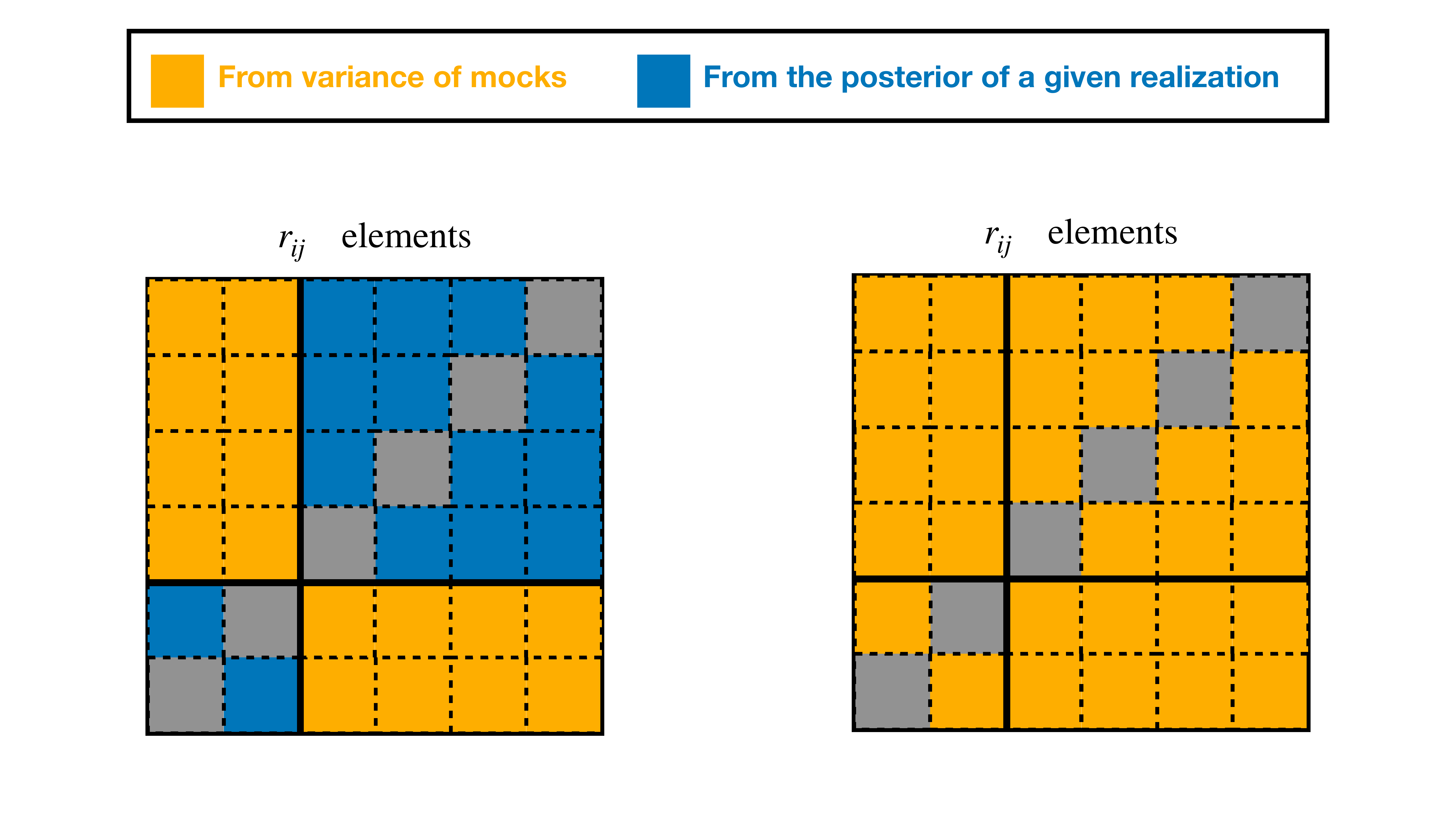}
    \includegraphics[scale=0.18]{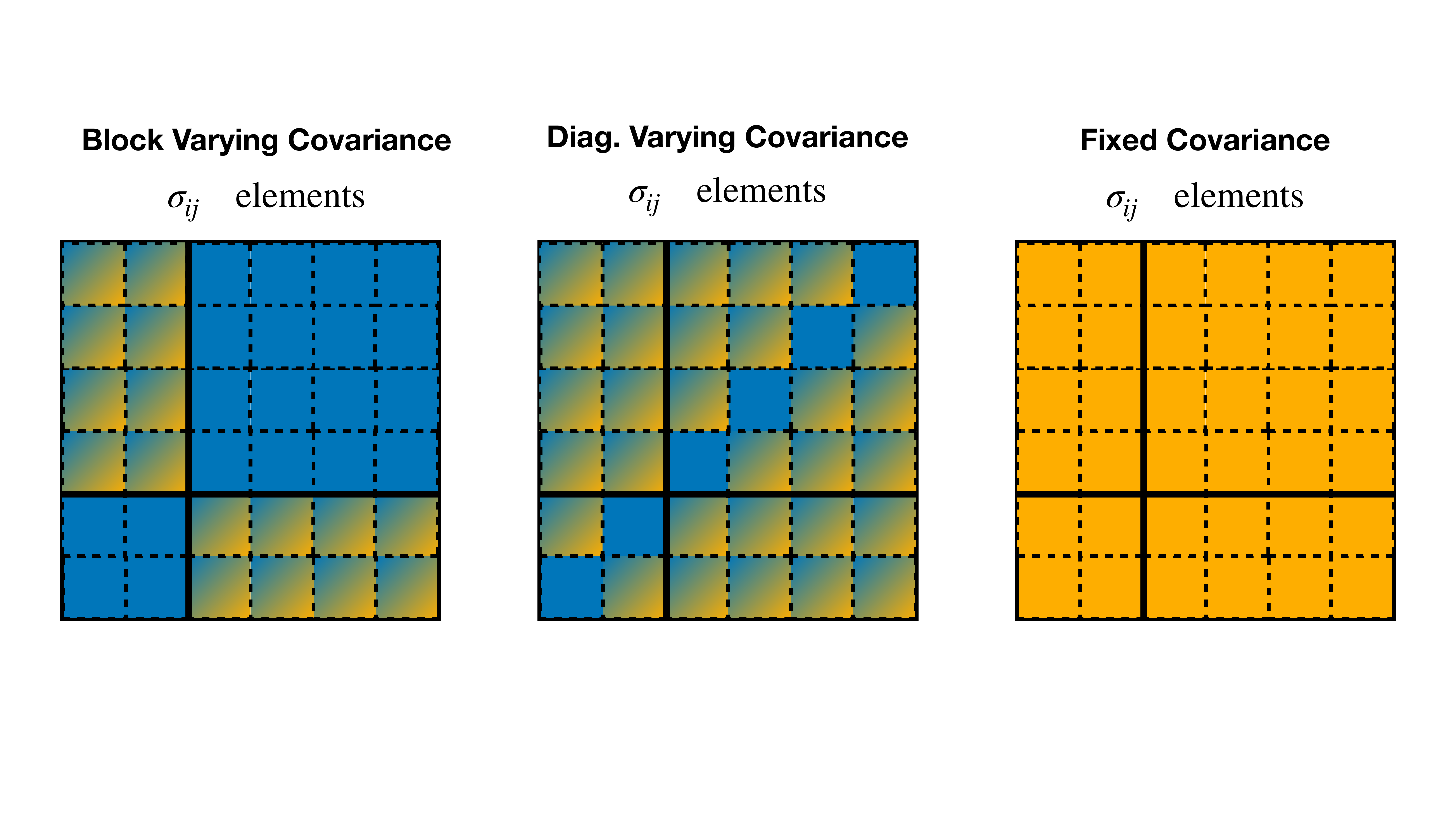}
     \caption{Schematic types (or architectures) of covariance matrices explored in this paper, the top row represents the cross-correlation coefficients, $r_{ij}$, and the bottom row the matrix elements, $\sigma_{ij}$ (see Eq. \ref{eq:matrix}). In this figure the $DD$ approach of FS-BAO covariance is displayed, with two elements for the BAO derived quantities, $D^{\rm BAO}=\{\alpha_\parallel,\alpha_\perp\}$, and four for full shape, $D^{\rm FS}=\{\alpha_\parallel,\alpha_\perp,\widetilde{f},m \}$. The solid lines divide the full covariance in the three block-covariance types: two diagonal full shape and BAO blocks, and one off-diagonal cross block. The `block varying' covariance architecture is built using the diagonal-block elements taken from the posteriors of each realization (blue squares). The off-diagonal block elements are constructed by combining the cross-correlation coefficients, $r_{ij}$, estimated from the variance of the mocks (yellow squares in the upper panels) with the diagonal elements, $\sigma_i$, from each posterior (resulting in the half blue, half yellow squares). Conversely, in the `fixed' covariance architecture, both the coefficients $r_{ij}$ and diagonal elements, $\sigma_i$, are estimated from the variance of the mocks, and consequently, are the same for all realizations. An intermediate architecture type is the `diagonal varying' covariance, where all the cross-correlation coefficients, $r_{ij}$, are computed from the mocks and the diagonal elements, $\sigma_i$, from the posteriors. Similarly, the same architecture types of covariances can be derived for the $PD$ approach. In this case only the BAO diagonal block is estimated from the posterior in both the block and `diagonal varying' covariance architectures.}
     \label{fig:covtype}
\end{figure}

For the $DD$ approach, the $\sigma_{ij}^{(m)}$ elements represent the covariance of the compressed variables. These elements can be estimated from the variance of mocks or from theory, as for the $PP$ approach, which would result in a unique realization-independent covariance (`fixed covariance' in Fig.~\ref{fig:covtype}). However, we know that when we display the covariance elements estimated from the posteriors of $D^{\rm BAO,\,FS}$ of each individual mock realization, these vary significantly among realizations. An example of this variation can be seen in fig. 8 of \cite{Gil-Marin:2020bct}, where the errors of $\alpha_{\parallel,\,\perp}$, placed as diagonal elements of the covariance, vary up to a factor of 2 for different realizations. It is clear then that treating the covariance of the compressed data-vector as fixed across independent realizations of the same sample might not be accurate. Unlike the variance of the power spectrum elements in the $PP$ approach, the variance of the compressed elements is not guaranteed to be invariant for different realizations. This is caused by the impact of the random phases of the initial condition field into the compressed variables variance. An alternative approach is to construct a different covariance for each realization (see the two `varying covariance' cases in Fig.~\ref{fig:covtype}), and thus account for the intrinsic variation with respect to realizations (or initial conditions of the field). We refer to the `block varying' covariance when the $\sigma^{(m)}_{ij}$ elements are computed by taking the diagonal-block elements from the posteriors of each realization.\footnote{Recall that for both BAO and FS diagonal blocks, the covariance elements have been estimated from a fixed $P$-covariance similarly to the $PP$-approach.} Then the off-diagonal block elements are computed as follows, the cross-correlation coefficients, $r^{\rm BAO-FS}_{ij}$ are derived from the variance of the mocks and therefore are fixed across realizations. The realization-dependence is reintroduced in the $\sigma_{ij}$ terms when the cross-correlation coefficient is multiplied by the diagonal elements, for e.g., $\sigma^{\rm BAO-FS}_{ij}=\sigma_{i}^{(\rm BAO)}\sigma_{j}^{(\rm FS)}r^{\rm BAO-FS}_{ij}$ (half yellow and blue squares in Fig.~\ref{fig:covtype}). Similarly, the `diagonal varying' covariance is constructed only taking the diagonal elements from the posteriors of each realization, and all the cross-correlation coefficients (including those of the diagonal blocks) from the variance of the mocks. Hence, all $r_{ij}$ are constant across realizations, but the $\sigma_{ij}$ elements are rescaled in a realization-dependent way via the $\sigma_i$ and $\sigma_j$ diagonal elements.

We refer to these three types of covariances, `fixed', `block varying' and `diagonal varying' as architectures, and are visually described in the bottom panels of Fig.~\ref{fig:covtype} for the $DD$ approach.
For the $PD$ approach three types of covariances can be equally drawn. In this case the diagonal block corresponding to the $P_{\rm pre}^{(\ell)}(k_i)$ elements is equal for all the mocks (`fixed', as for the $PP$ approach), and only the diagonal $D^{\rm BAO}$ block, and the off-diagonal block elements, are susceptible to be estimated by either the `fixed', `block varying' or `diagonal varying' approaches, just as for the $DD$ case, {\it mutatis mutandis}.

For the main analysis of this paper we choose the `block varying' covariance case, for both $PD$ and $DD$ approaches. Later in  \S\ref{sec:sys} we will explore the impact that the different covariance estimators of Fig.~\ref{fig:covtype} have in the distribution of errors for the $PD$ and $DD$ approaches, which will justify this choice. Note that for the $PP$ approach we always use the `fixed' architecture, which is supported by our understanding of the statistical properties of cosmic structures.

\section{Results}\label{sec:results}

\begin{figure}[ht]
\centering
    \includegraphics[scale=0.5]{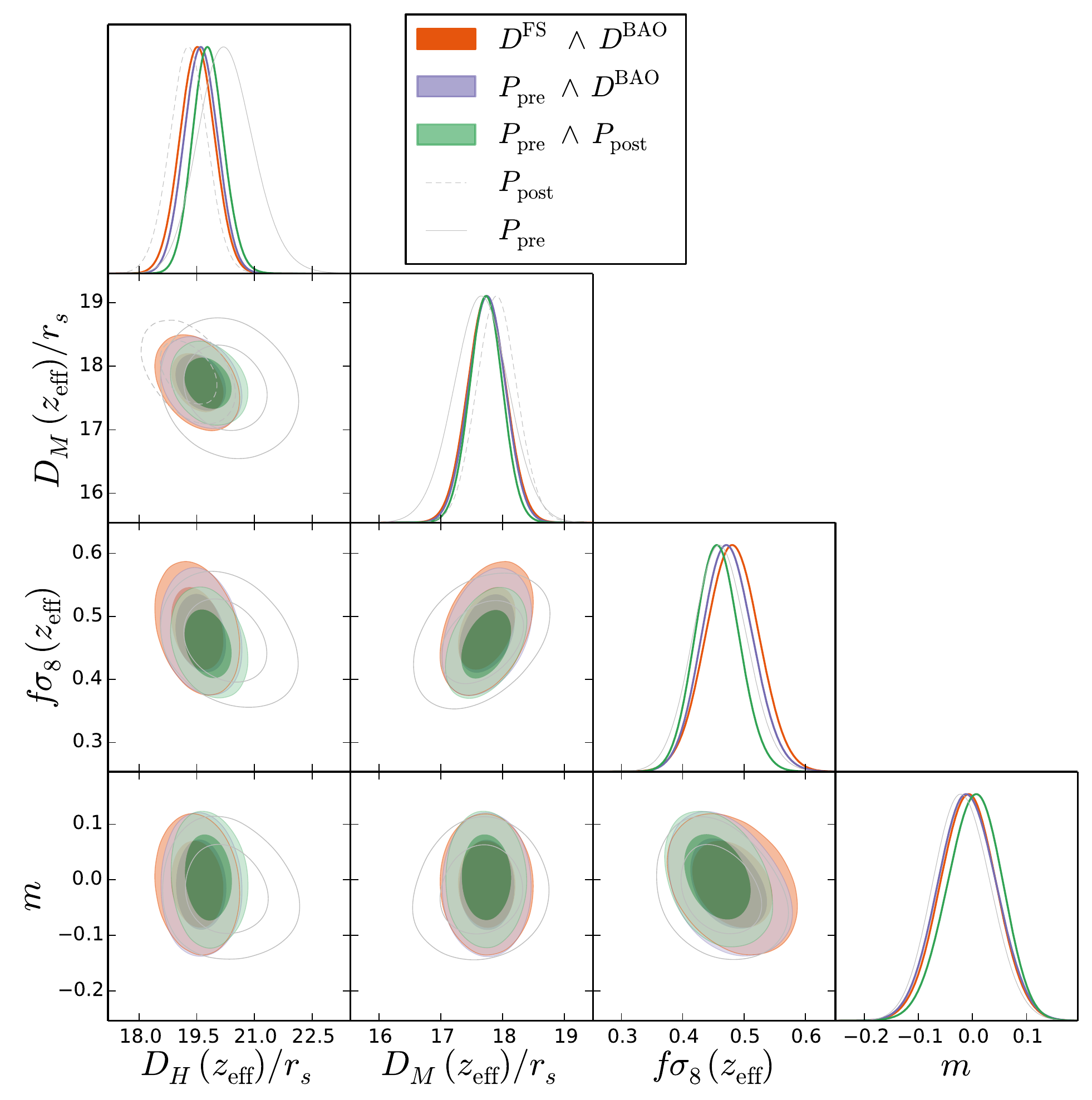}
        \caption{Posteriors of the four cosmological variables of the combined data-vector, $\{\alpha_\parallel,\,\alpha_\perp,\,\widetilde{f},\,m\}$, (converted into the physical variables $\{D_H/r_s,\,D_M/r_s,\,f\sigma_8,\,m\}$) corresponding to the eBOSS LRG data catalogues at $z_{\rm eff}=0.70$. The different colors display the three different methodologies described in  \S\ref{sec:methodology} to combine pre- and post-recon catalogues, as written in the key. For reference, the posteriors corresponding to the analysis of only pre- and post-recon catalogues are also shown in solid black and dashed black lines, respectively. The numerical results of this figure are displayed in Table~\ref{table:data}.}
        \label{fig:data}
\end{figure}

We start by running the three approaches, $PP$, $PD$ and $DD$, on the eBOSS LRG data catalogue. The posteriors for the $D^{\rm FS+BAO}$ data-vector are displayed in Fig.~\ref{fig:data} in different colors, accordingly. In addition, we show in black lines the posteriors when only the pre- (solid lines) and post-recon (dashed lines) catalogues are used. 
The agreement among the three approaches is remarkable. We note that for all variables the $PP$ approach tends to have tighter posteriors, suggesting that this approach may have more constraining power than $DD$ or $PD$. This is expected, as combining the summary statistics of pre- and post-recon catalogues can introduce non-Gaussian noise. This is for example the case when the BAO is not detected when analyzing the pre-recon data-vector, but it is for the analysis of the post-recon. The non-BAO detection makes the shape of the pre-recon BAO posteriors to be non-Gaussian, with long tails that usually can even hit the prior limits. This non-Gaussian behaviour can be solved if the poorly-constrained BAO variables of the pre-recon data-vector are anchored by the BAO signal in the post-recon data-vector, as it happens when simultaneously fitting both (for the $PP$ approach). The variables most affected by this choice of approaches are $D_H/r_s$ and $f\sigma_8$. Again, this behaviour is expected. The longitudinal BAO is detected at lower signal-to-noise than the transverse BAO because of the density of $k$-modes contributing: $N_k^2$ for the transverse BAO and $N_k$ for the longitudinal BAO, where $N_k$ is the number of modes in a given $k$-bin. On the other hand, the RSD signal is degenerated with the anisotropy generated by the Alcock-Paczynski effect (mainly driven by the detection of BAO), and therefore, the better the BAO is resolved, the better $f\sigma_8$ is measured. Table~\ref{table:data} displays the numerical results corresponding to Fig.~\ref{fig:data}.

\begin{table}[ht]
    \centering
    \begin{tabular}{|c|c|c|c||c|c|}
    \hline
        variable & $P_{\rm pre}\wedge P_{\rm post}$ & $P_{\rm pre}\wedge D^{\rm BAO}$ & $D^{\rm FS}\wedge D^{\rm BAO}$ & $P_{\rm pre}$ & $P_{\rm post}$\\
        \hline
        \hline
        $D_H/r_s$ & $19.83\pm 0.41$ & $19.63\pm0.43$ & $19.54 \pm 0.45$ & $20.26\pm0.72$ & $19.31\pm0.49$\\
        $D_M/r_s$ & $17.69\pm 0.27$ & $17.71\pm 0.29$ & $17.70 \pm 0.31$ & $17.60\pm0.44$ & $17.87\pm0.33$\\
        $f\sigma_8$ & $0.455\pm 0.036$ & $0.472 \pm 0.041$ & $0.479 \pm 0.043$ & $0.461\pm0.044$ & -\\
        $m$ & $0.003\pm 0.051$ & $-0.010 \pm 0.053$ & $-0.008\pm0.052$ & $-0.019\pm0.053$ & -\\
        \hline
    \end{tabular}
    \caption{Numerical results from combining the pre- and post-recon measurements for the eBOSS LRG data catalogue. The different columns display the results corresponding to the three different approaches described in  \S\ref{sec:methodology} to combine pre- and post-recon catalogues. The results for the pre- and post-recon analyses alone are added for completeness. These values correspond to the results displayed in Fig.~\ref{fig:data}.}
    \label{table:data}
\end{table}

\begin{figure}[ht]
    \centering
    \includegraphics[scale=0.35]{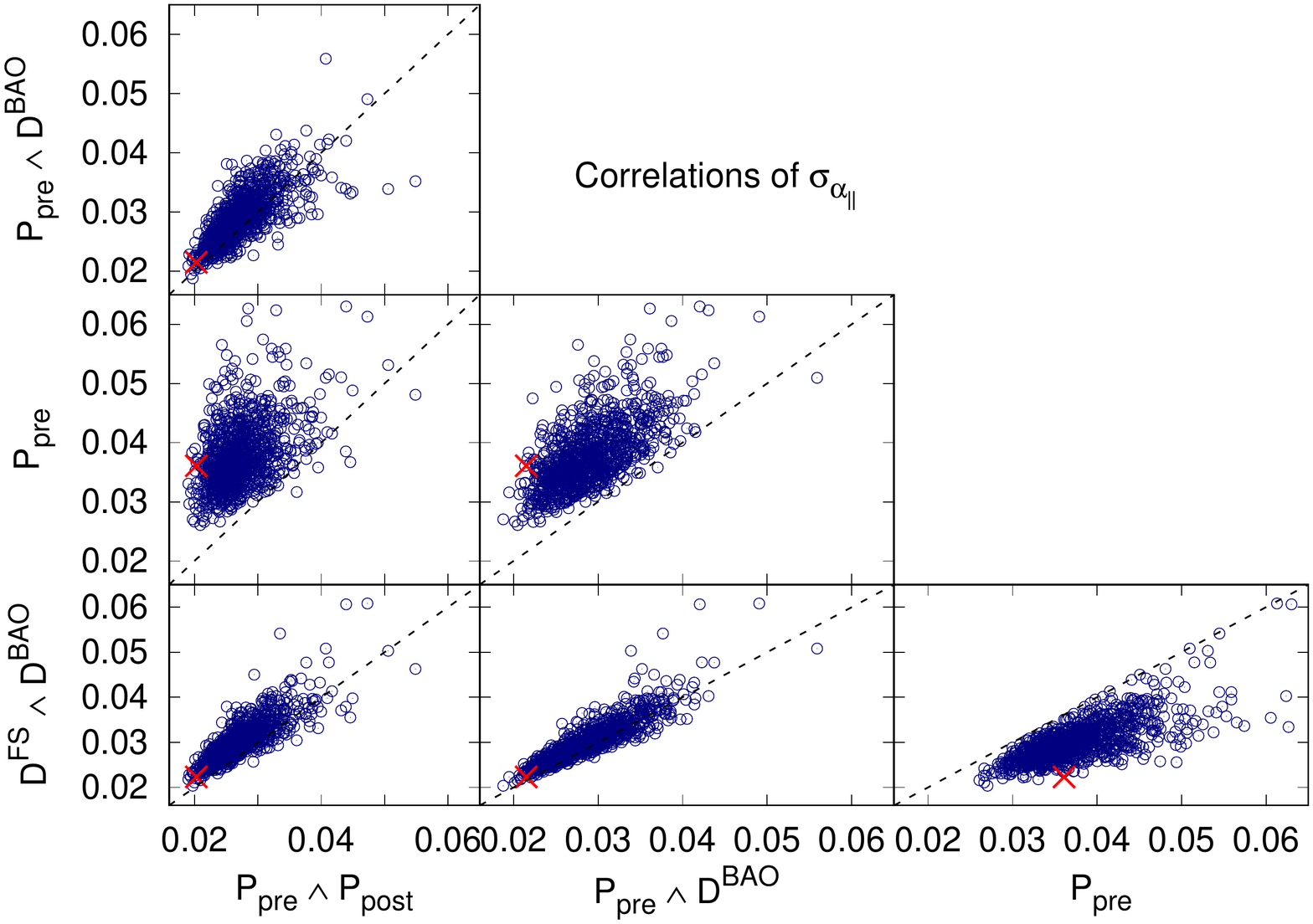}
    \includegraphics[scale=0.35]{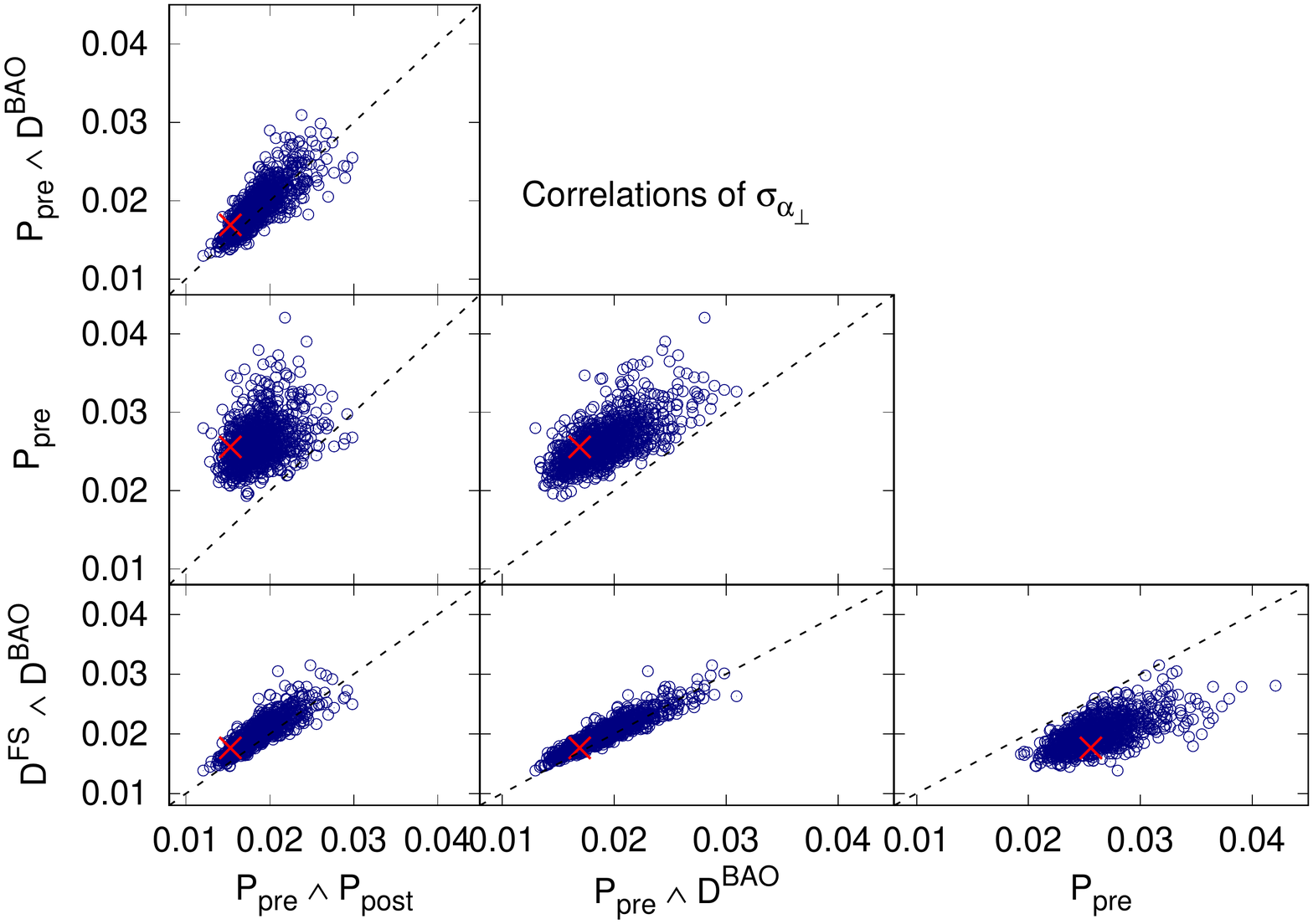}
    \caption{Error values of $\alpha_\parallel$ ($\sigma_{\alpha_\parallel}$; top panel) and $\alpha_\perp$ ($\sigma_{\alpha_\perp}$; bottom panel) for the 1000 EZ-mocks (blue dots) and for the data (red-cross) of the eBOSS LRG catalogues. For each case the results are shown for the three approaches described in \S\ref{sec:methodology}: summary statistics $P_{\rm pre}\wedge P_{\rm post}$ ($PP$ approach); compressed variables $D^{\rm FS}\wedge D^{\rm BAO}$ ($DD$ approach); and the hybrid $P_{\rm pre}\wedge D^{\rm BAO}$ ($PD$ approach). The case when only the pre-recon catalogue is used is also shown for reference ($P_{\rm pre}$). The overall distribution of errors is very consistent among the three studied cases and lies along the diagonal dashed line for all the sub-panels (except for the pre-recon case, where the errors are larger compared to the rest of the approaches).}
    \label{fig:corr}
\end{figure}

However, the results shown for the data correspond just to a single realization, which may not be representative of the typical behaviour. For this reason, we repeat this exercise on each realization of the 1000 mocks. The results are shown for the errors of $\alpha_\parallel$ and $\alpha_\perp$ in Fig.~\ref{fig:corr}, in the top and bottom panels, respectively. The blue dots represent the errors for each mock realization, and the red cross for the data, just for reference. The corresponding plot of the actual values of $\alpha_{\parallel,\,\perp}$ (and not their errors) can be found in Appendix \ref{sec:alphas}, along with the expected values for the mocks. 

We observe that these three cases also look very consistent across all the independent realizations of the mocks, as the errors for the different realizations lie along the diagonal dashed line in the sub-panels. In addition, we confirm the performance of the data in terms of error-bars in Fig.~\ref{fig:data} is typical  when compared to the mocks realizations. As expected, the cloud of points is more spread in the distribution of errors of $\sigma_{\alpha_\parallel}$ than in the $\sigma_{\alpha_\perp}$, due to the number of $N_k$ modes contributing, as discussed above. 

We also see that when adding the reconstructed catalogue into the analysis, the error of $\alpha_{\parallel,\,\perp}$ is reduced with respect to the pre-recon analysis case ($P_{\rm pre}$), as expected. This happens for all of the three approaches considered. We also see that the effect of $PP$ reporting a smaller error for the data on the $\alpha$'s is a general trend for most of the mocks, as the cloud of points tend to be slightly above the diagonal line in the left-hand-side column of sub-panels, in both top and bottom panels of Fig.~\ref{fig:corr}. This effect can be seen more clearly in Table~\ref{tab:mocks}, where different statistics are shown for $\alpha_\parallel$, $\alpha_\perp$ and $\widetilde{f}$ variables computed from the 1000 mocks.\footnote{Although computed, we do not display the results for the variable $m$ for conciseness. As shown in Fig.~\ref{fig:data} and Table~\ref{table:data}, this is the least affected variable by the choice of combining catalogues.} 

\begin{table}[ht]
    \centering
    \begin{tabular}{|c|cc|c||cc|c||cc|c|}
    \hline
           compression & $\langle \sigma_{\alpha_\parallel} \rangle$ & $S_{\alpha_\parallel}$  & $S_{Z_{\alpha_\parallel}}$  & $\langle \sigma_{\alpha_\perp} \rangle$  & $S_{\alpha_\perp}$  & $S_{Z_{\alpha_\perp}}$  &  $\langle\sigma_{\widetilde{f}}\rangle$ & $S_{\widetilde{f}}$ & $S_{Z_{\widetilde{f}}}$ \\
         \hline
         \hline
         $P_{\rm pre}$ & 3.776 & 3.872 & 1.018 & 2.617 & 2.481 & 0.940 & 8.795 & 8.386 & 0.949 \\
         $P_{\rm post}$ & 4.056 & 3.966 & 0.923 & 2.386 & 2.348 & 0.967 & - & - & - \\
         \hline
         $D^{\rm FS} \wedge D^{\rm BAO}$ & 3.023 & 3.146 & 1.005 & 2.000 & 1.936 & 0.967 & 8.551 & 8.167 & 0.954 \\
         $P_{\rm pre}\,\wedge\, D^{\rm BAO}$ & 2.890  & 2.986  & 1.026  &  1.917 & 1.813  & 0.952  & 8.349 & 7.865 & 0.941 \\
         $P_{\rm pre}\,\wedge\,P_{\rm post}$ & 2.727 & 2.824 & 1.015 & 1.882 & 1.825 & 0.981 & 7.801 & 7.232 & 0.926 \\
         \hline
    \end{tabular}
    \caption{Statistical quantities drawn from the 1000 realizations of the mocks. The $\langle \sigma_x\rangle$ display the average of the errors of $x$, whereas $S_x$ display the {\it rms} of the variable $x$. The $S_{Z_x}$ displays the {\it rms} of $Z_x\equiv (x-\langle x\rangle)/\sigma_x$. We are only displaying the results for $x=\alpha_\parallel,\,\alpha_\perp,\,\widetilde{f}$, although $m$ and several nuisance parameters are also varied in the analysis. For all cases the values of $\sigma_x$ and $S_x$ are multiplied by $100$ for a better reading.}
    \label{tab:mocks}
\end{table}

In Table~\ref{tab:mocks}, the $\langle \sigma_x\rangle$ represents the average of the errors of the variable $x$; the $S_x$ displays the {\it rms} of the variable $x$; the $S_{Z_x}$ displays the {\it rms} of the variable $Z_{x}$, defined as, $Z_x\equiv (x-\langle x \rangle)/\sigma_x$. Both $\langle \sigma_x\rangle$ and $S_x$ are estimators of the error of $x$. In the ideal case of Gaussian distributions, where the covariance estimators are unbiased (i.e., the errors of the data-vector are not over- or under-estimated), both error estimators, $\langle \sigma_x\rangle$ and $S_x$, should report the same value (within certain numerical noise). In this fashion, $S_{Z_x}$ would be very close to unity. Deviations from unity, for e.g., $S_{Z_x}>1$, are a smoking gun on the underestimation of the $\sigma_x$ (or overestimation for $S_{Z_x}<1$). We need to bear in mind that this holds only if the distributions of $\sigma_x$ and $x$ are Gaussian. For non-Gaussian distribution the values of $S_{Z_x}$ can be either higher or lower than unity for a non-biased errors, depending on the amount of skewness and kurtosis that the distribution of $x$ and $\sigma_x$ may have. 

From Table \ref{tab:mocks} we see that the $\langle \sigma_x\rangle$ values for $x=\alpha_\parallel$, $\alpha_\perp$ and $\widetilde{f}$ are $5-10\%$ smaller for the $P_{\rm pre}\wedge P_{\rm post}$ than for the other two cases, as seen for the data in Table~\ref{table:data}. Also, we check that the same applies for the $S_x$ variable, which is insensitive to scale-independent biases in the covariance matrix, and hence more insensitive to potential inaccuracies present in the covariance. In general, all the three approaches have the same degree of deviation from $S_{Z_x}=1$, which is always significantly less than $10\%$: $1-2\%$ for $\alpha_\parallel$ towards the under-estimation; $8-5\%$ for $\alpha_\perp$ and $\widetilde{f}$ towards the over-estimation. These small discrepancies can be due to small non-Gaussian components in the statistical distribution of the quantities and their errors. In Table \ref{tab:moments} the third (skewness) and fourth (kurtosis) statistical moments of the distribution of $\alpha_\parallel$ and $\alpha_\perp$ are shown, for the cases studied in Table \ref{tab:mocks}. Certainly, we see some deviations from the expected values of a Gaussian distribution, 0 for the skewness and 3 for the kurtosis, where the power spectrum inferred variables ($PP$) display a more Gaussian distribution than the compressed-inferred variables ($DD$). These results suggest that when building a covariance matrix as a Gaussian approximation for the $D^{\rm FS}$ and $D^{\rm BAO}$ data-vectors ($DD$ approach) we end up introducing some non-Gaussian noise compared to taking the full covariance matrix from the power spectrum bins ($PP$ approach). Also, the $\alpha_\parallel$ statistical moments tend to deviate more from the Gaussian prediction than the $\alpha_\perp$ ones, but probably just because it has intrinsically more noise (the BAO is worse measured along ($\alpha_\parallel)$ than across the line-of-sight ($\alpha_\perp$)). 

\begin{figure}[ht]
    \centering
    \includegraphics[scale=0.21]{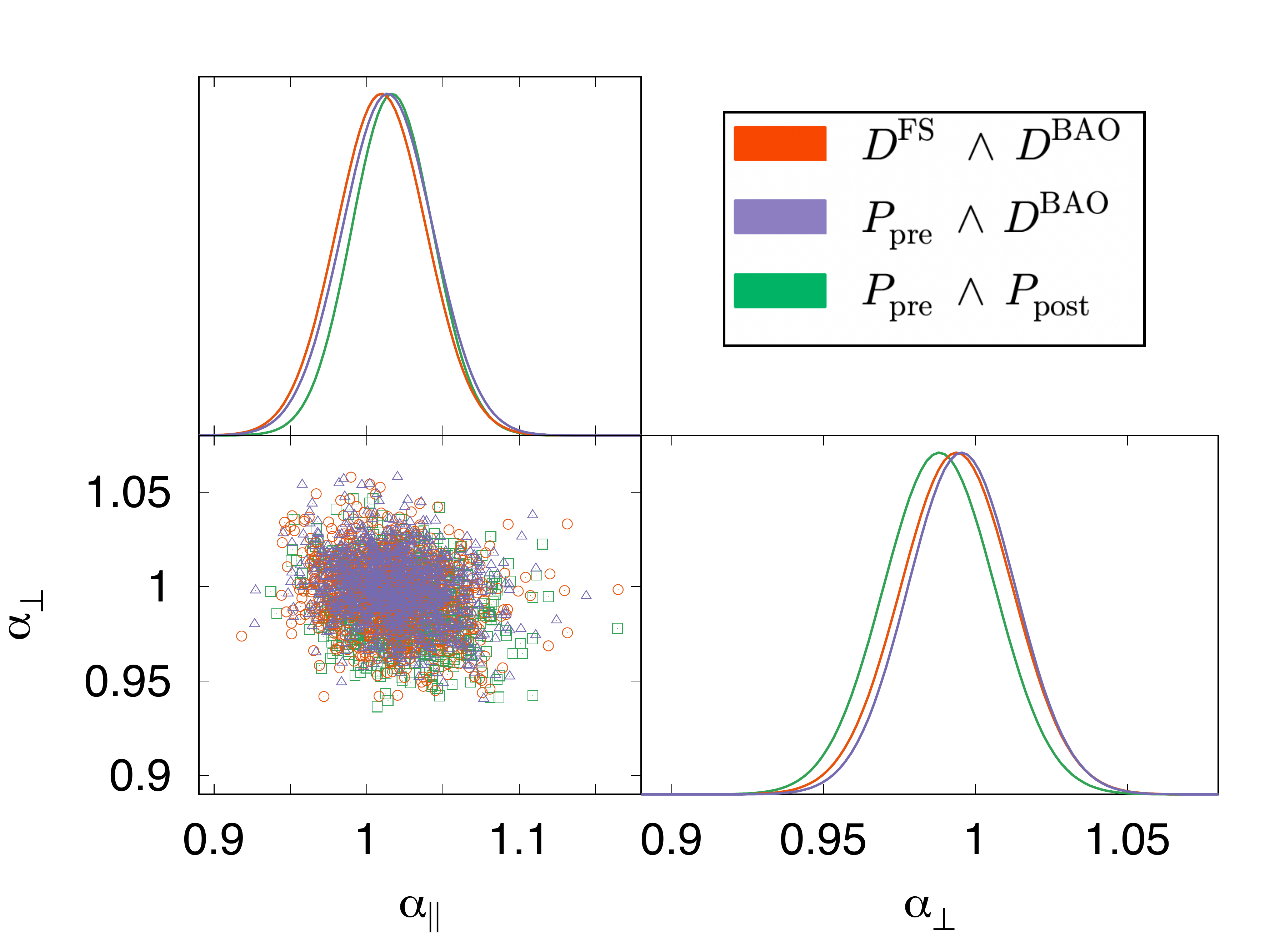}
    \includegraphics[scale=0.21]{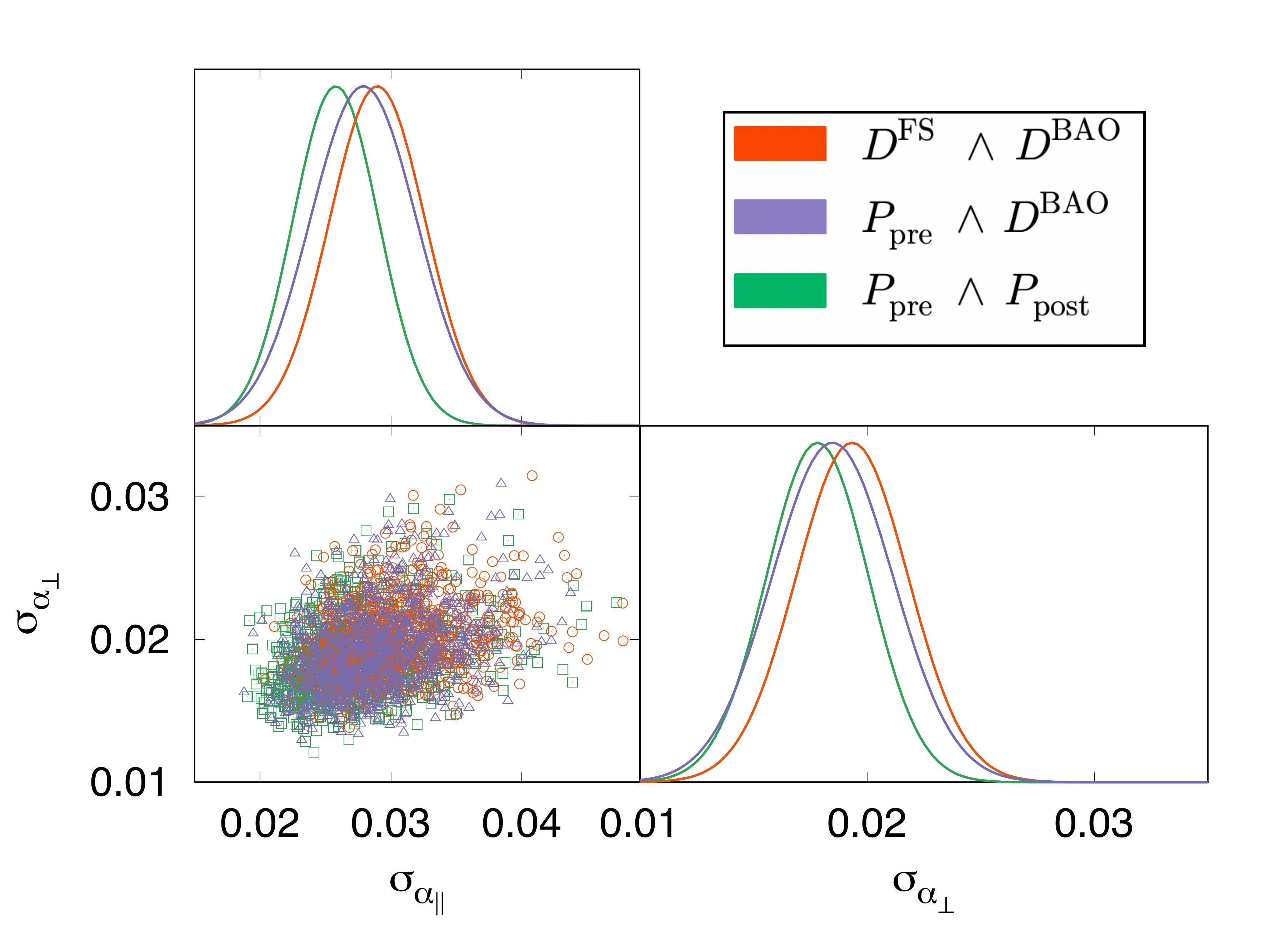}
    \caption{Distribution of the measured $\{\alpha_\parallel,\,\alpha_\perp\}$ (left panel) and their errors (right panel), for the $PP$ approach (green), the $PD$ approach (purple) and the $DD$ approach (orange), using the same color coding as in Fig.~\ref{fig:data}. The 1D posteriors show the Gaussian approximation to the measured histogram.}
    \label{fig:histograms}
\end{figure}

Fig.~\ref{fig:histograms} displays the distribution of $\alpha_\parallel-\alpha_\perp$ (left panel) and $\sigma_{\alpha_{\parallel}}-\sigma_{\alpha_\perp}$ using the same color coding as in Fig.~\ref{fig:data}. The 1D posteriors are the Gaussian approximation from the cloud of points. As before we see that the width of the distribution of the $\alpha$'s is slightly smaller for the $PP$ case in the left panel, and the center of the distribution of the $\sigma_{\alpha's}$ has a smaller value for the $PP$ case, just in line what the reported values of Table~\ref{tab:mocks}. From the left panel of Fig.~\ref{fig:histograms} we observe a slight difference in the peak of the Gaussian approximation for the $PP$ approach compared to those of the $PD$ and $DD$ approaches. This offset is quantified in Appendix \ref{sec:alphas}.

\begin{table}[ht]
    \centering
    \begin{tabular}{|c||c|c||c|c|}
         \hline
        compression &  $\alpha_\parallel$ skewness & $\alpha_\parallel$ kurtosis &  $\alpha_\perp$ skewness & $\alpha_\perp$ kurtosis  \\
         \hline
         \hline
         $P_{\rm pre}$ & $0.346\pm0.156$ & $3.35\pm0.21$ & $-0.016\pm0.074$ & $3.30\pm0.12$ \\
         $P_{\rm post}$ & $0.029\pm0.202$ & $5.77\pm0.62$ & $0.041\pm0.117$ & $3.71\pm0.23$\\
         \hline
         $D^{\rm FS} \wedge D^{\rm BAO}$  & $0.764\pm0.143$ & $6.18\pm 0.94$ & $-0.291\pm0.100$ & $4.43\pm 0.49$ \\
         $P_{\rm pre}\,\wedge\, D^{\rm BAO}$  & $0.261\pm 0.076$ & $3.25\pm0.20$ & $0.042\pm0.098$ & $3.18\pm0.15$ \\
         $P_{\rm pre}\,\wedge\,P_{\rm post}$  & $0.383\pm0.094$ & $3.77\pm0.50$ & $-0.054\pm0.054$ & $2.95\pm0.12$ \\
         \hline
    \end{tabular}
    \caption{High-order statistical moments for $\alpha_\parallel$ and $\alpha_\perp$ drawn from 1000 realizations of the mocks, estimated using the three methods described in  \S\ref{sec:methodology}. For reference also the derived quantities for the pre- and post-recon catalogues are shown. For a pure Gaussian distribution we expect a skewness of 0 and a kurtosis of 3. The errorbars represent the {\it rms} obtained when the 1000 mocks are split in 10 chunks of 100 mocks, appropriately rescaled for the whole set of 1000 mocks. }
    \label{tab:moments}
\end{table}

In summary, we find that the $PP$ approach presents slightly smaller errors on the compressed variables, $\alpha_\parallel$, $\alpha_\perp$ and $f\sigma_8$ than the other two approaches, which result in a $5-10\%$ more constraining power. Furthermore, the resulting distribution from the $PP$ approach on the $D^{\rm FS+BAO}$ variables better follows a Gaussian statistic than the $DD$ approach, which makes the statistical interpretation easier. This trend has been checked not only for the data, but also for the set of mocks, hence  is a common feature independently on the initial conditions of the sample. 

\section{Systematic checks}\label{sec:sys}

In this section we explore a set of potential systematic effects that may impact the results presented in \S\ref{sec:results}. In particular, we focus on the architecture of the covariance for the $DD$ and $PD$ approaches; the number of mocks used for estimating the covariance for the $PP$ approach; and the broadband polynomial order used in the modelling when doing a combined BAO and full-shape type of analysis under the $PP$ approach. 

\subsection{Architecture of the covariance matrix}

We aim to test the effect of the different types of covariance matrix architectures for the $DD$ and $PD$ approaches. These different architectures have been already summarized in Fig.~\ref{fig:covtype}. In the previous section, we have decided to take as the fiducial approach the `block varying' covariance. As described in \S\ref{sec:methodology} this type of covariance architecture applied to the $DD$ approach takes the diagonal blocks from the independent full shape and BAO analyses (i.e., from the individual likelihoods of each mock). This means that these covariance blocks vary across realizations. For the $PD$ approach, this only applies to the BAO diagonal block, as the other diagonal block (the one which describes the covariance of the pre-recon power spectra) is kept fixed for all the realizations (as for the $PP$ approach), and the off-diagonal blocks also change for each realization.

\begin{figure}[htb]
    \centering
    \includegraphics[scale=0.27]{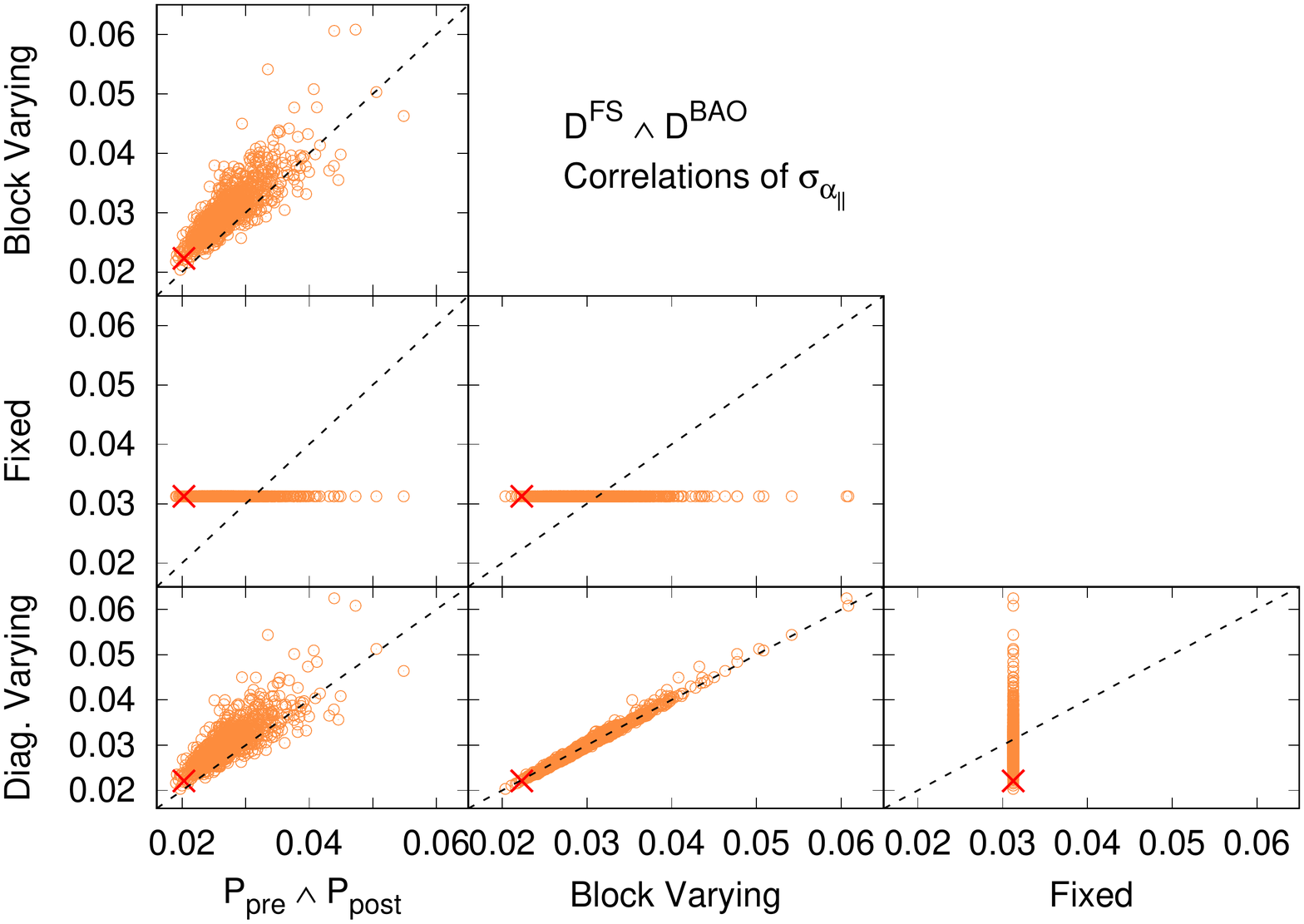}
    \includegraphics[scale=0.27]{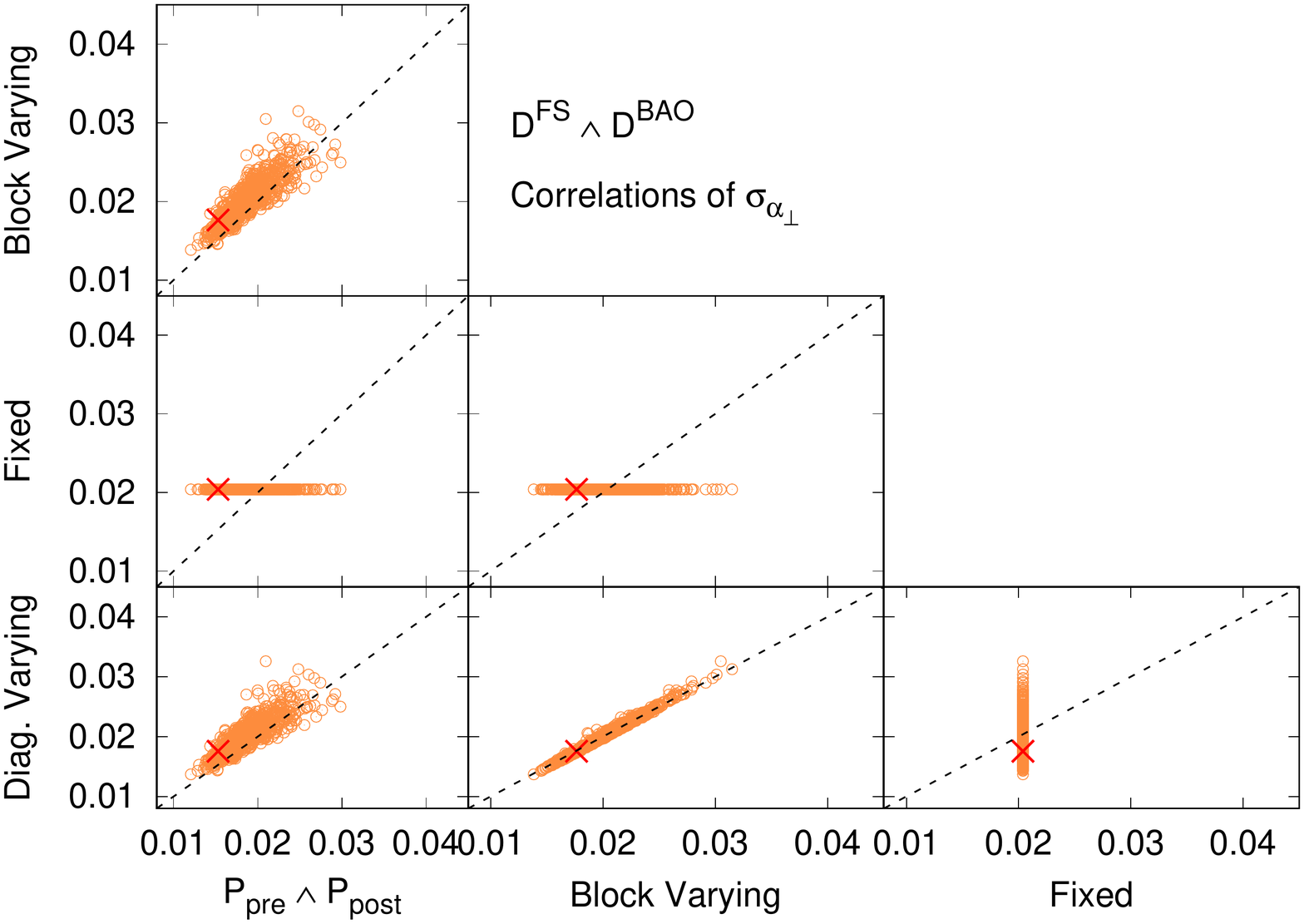}
    
        \includegraphics[scale=0.27]{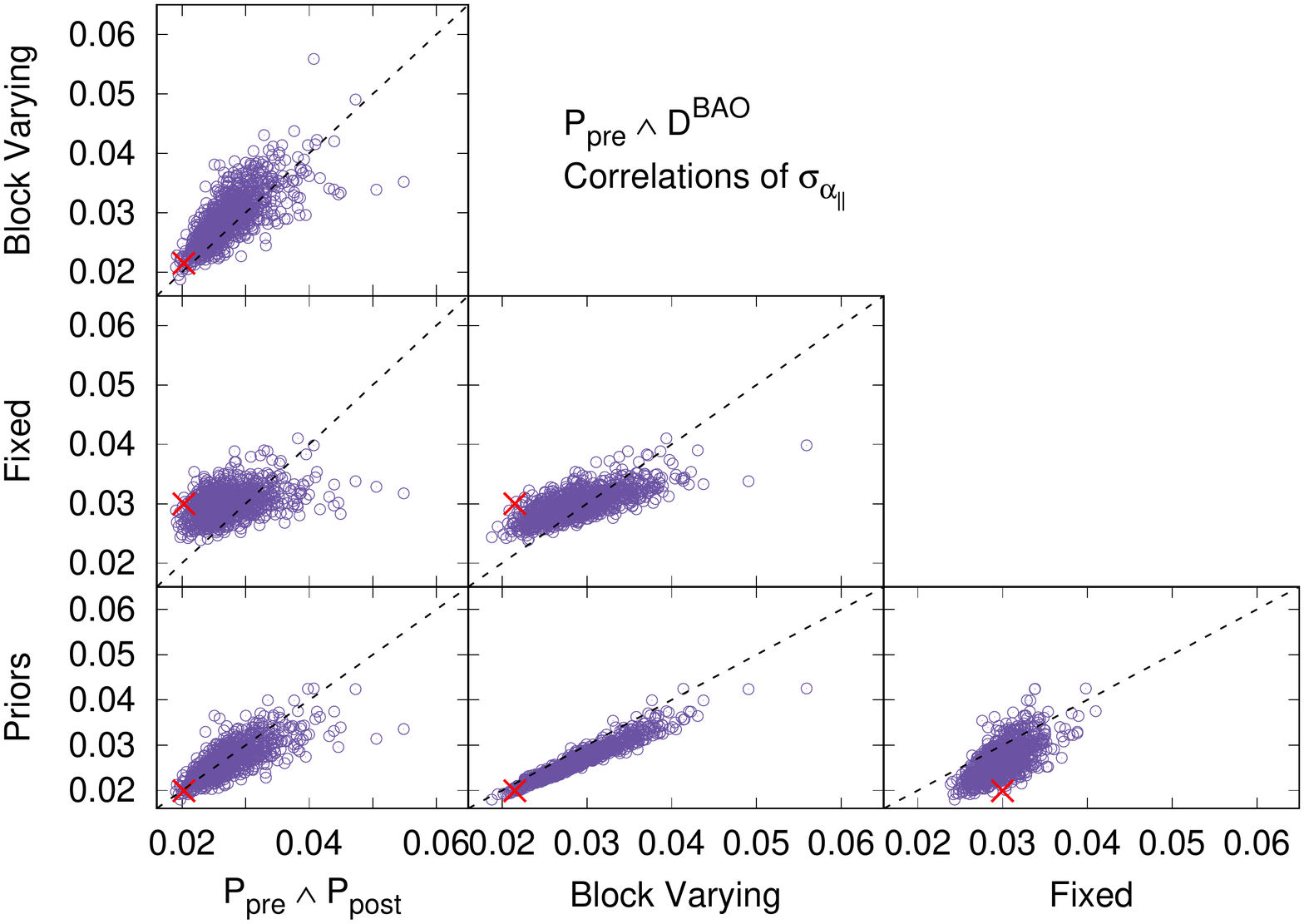}
    \includegraphics[scale=0.27]{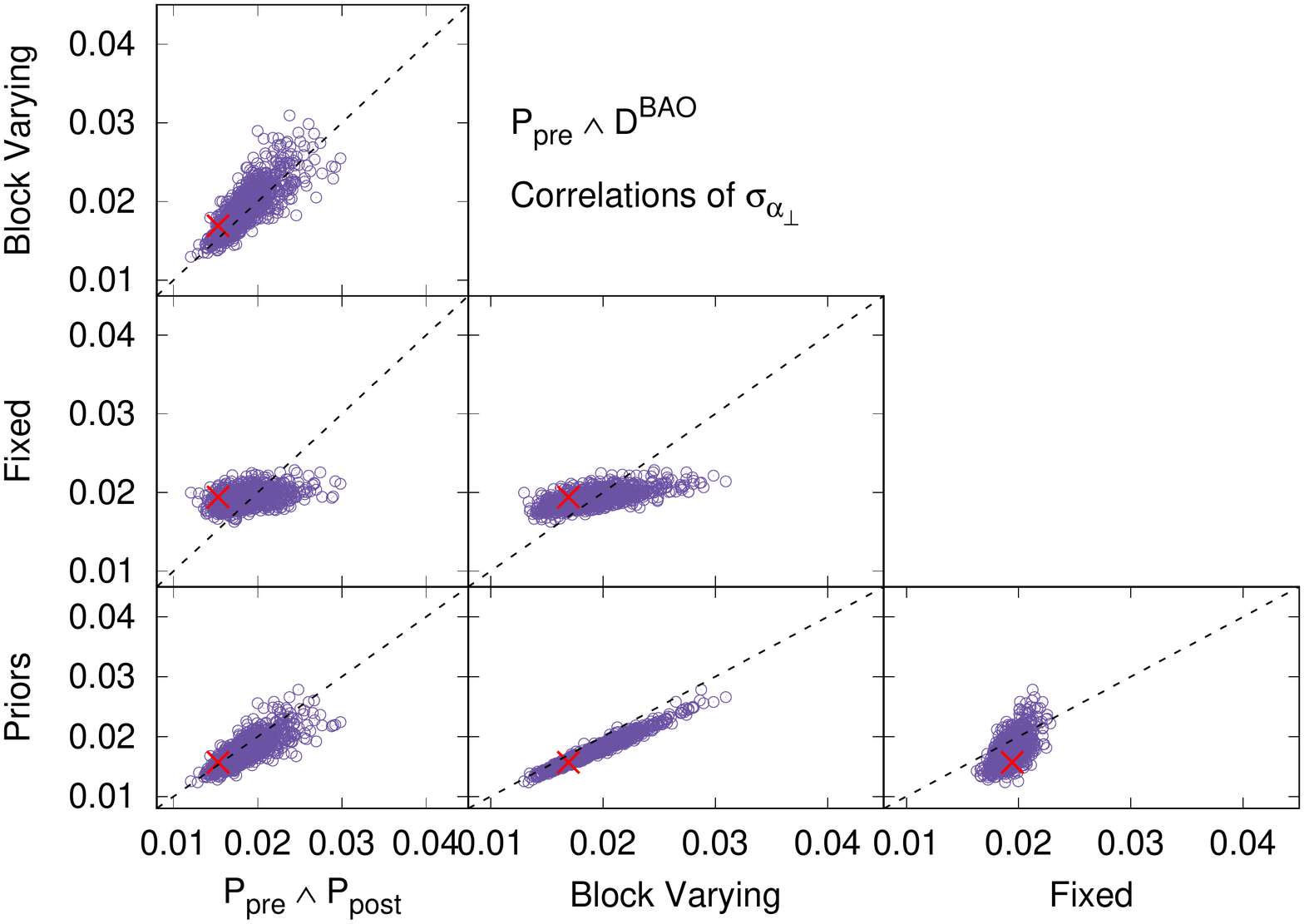}
    \caption{Effect of the different covariance architectures (described in Fig.~\ref{fig:covtype}) when following the $DD$ approach (top panels) and the $PD$ approach (bottom panels); for the errors of $\alpha_\parallel$ (left panels), and the errors of $\alpha_\perp$ (right panels). Each panel shows the results on the 1000 mocks (orange and purple circles, for $DD$ and $PD$, respectively) and the data (red cross). For each approach, the `block varying' and `fixed' covariance types are displayed. For the $DD$ also the `diagonal varying' case is tested. For the $PD$ also the varying prior case is tested (see text for details). The $PP$ approach  performance is shown in both cases as the reference of the expected results. This approach uses a `fixed' covariance in terms of the power spectra, which is the correct approach given the way the initial conditions are generated for the mocks, with a constant variance and random phases of the field.}
    \label{fig:FSpriors}
\end{figure}

The top panels of Fig.~\ref{fig:FSpriors} display the performance of the three types of covariances: `fixed', `block varying' and `diagonal varying', for the $DD$ approach. We also display the results from the $PP$ approach. This approach serves as a reference of the true and expected behaviour for the other two cases, as it relies on the assumption that variance of the field is the same for all realizations. The `block varying' covariance and the `diagonal varying' covariance architectures yield almost identical errors for $\alpha_\parallel$ (top left triangle plot) and $\alpha_\perp$ (top right triangle plot). Hence, these two types of architecture are almost equivalent. However, notice the striking difference in performance between the `fixed' and both `varying' (block and diagonal) covariances. The `fixed' covariance produces the same error on the parameters for the different mocks, independently of their realization, sometimes overestimating, and sometimes underestimating the errors with respect to the varying covariance cases. Note that when taking the average among all realizations the error is not significantly over or underestimated, but on a single realization it can be (and it is for most of the cases). For example, the error estimated for the data (red cross) happens to be larger for the `fixed' covariance case than for the varying covariance cases. Also, when we compare the $DD$ approach, implemented with the varying type of covariances, with the $PP$ approach, which has the right covariance (and errors), motivated by how the mocks have been generated, we find an excellent correspondence. This demonstrates that any of the choices based on the `varying covariances' and $DD$ yield the correct estimate of errors. For this reason we took the `block varying' diagonal as the fiducial case in \S\ref{sec:results}.

The bottom panels of Fig.~\ref{fig:FSpriors} display the same but for the $PD$ case. We have not displayed the `diagonal varying' covariance case because, as it happens for the top panels' it is virtually indistinguishable from the `block varying' architecture.  Instead, we show a complementary approach which we refer as `priors' (or `varying priors'), imposing  independent and uncorrelated priors on $\alpha_\parallel$ and $\alpha_\perp$ provided by the realization-specific posteriors for the BAO-only post-recon analyses. When comparing the `fixed' and the `block varying' covariances, we find a very similar performance as in the top panels. However, unlike for $DD$, for the $PD$ case considering the `fixed' covariance architecture only makes part of the covariance inaccurate, as the $P^{(\ell)}_{\rm pre}(k)-P^{(\ell')}_{\rm pre}(k')$ covariance block remains identical to the $PP$ case, and hence correct. Consequently, the behaviour is less extreme than in the $DD$ approach, and we observe some variability in the errors, although it is smaller than the one reported for the `block varying' case (or the $PP$ case). 
As for the $DD$ case, the `fixed' covariance in the $PD$ case tends to systematically over- or under-estimate the errors compared to the `block varying' cases, but keeping their average unbiased. The `varying prior' case reproduces quite closely across all realizations the error estimate of the `block varying' case. This implies that the off-diagonal blocks in the $PD$ covariance do not have a large impact in the final results. However, when ignoring them (as in the varying prior case) the resulting errors tend to be slightly under-estimated compared to the case where they are fully considered. Finally, as in the top plots, we find an excellent correspondence between the `block varying' covariance and the reference and correct $PP$ approach. 

\begin{figure}[ht]
    \centering
    \includegraphics[scale=0.5]{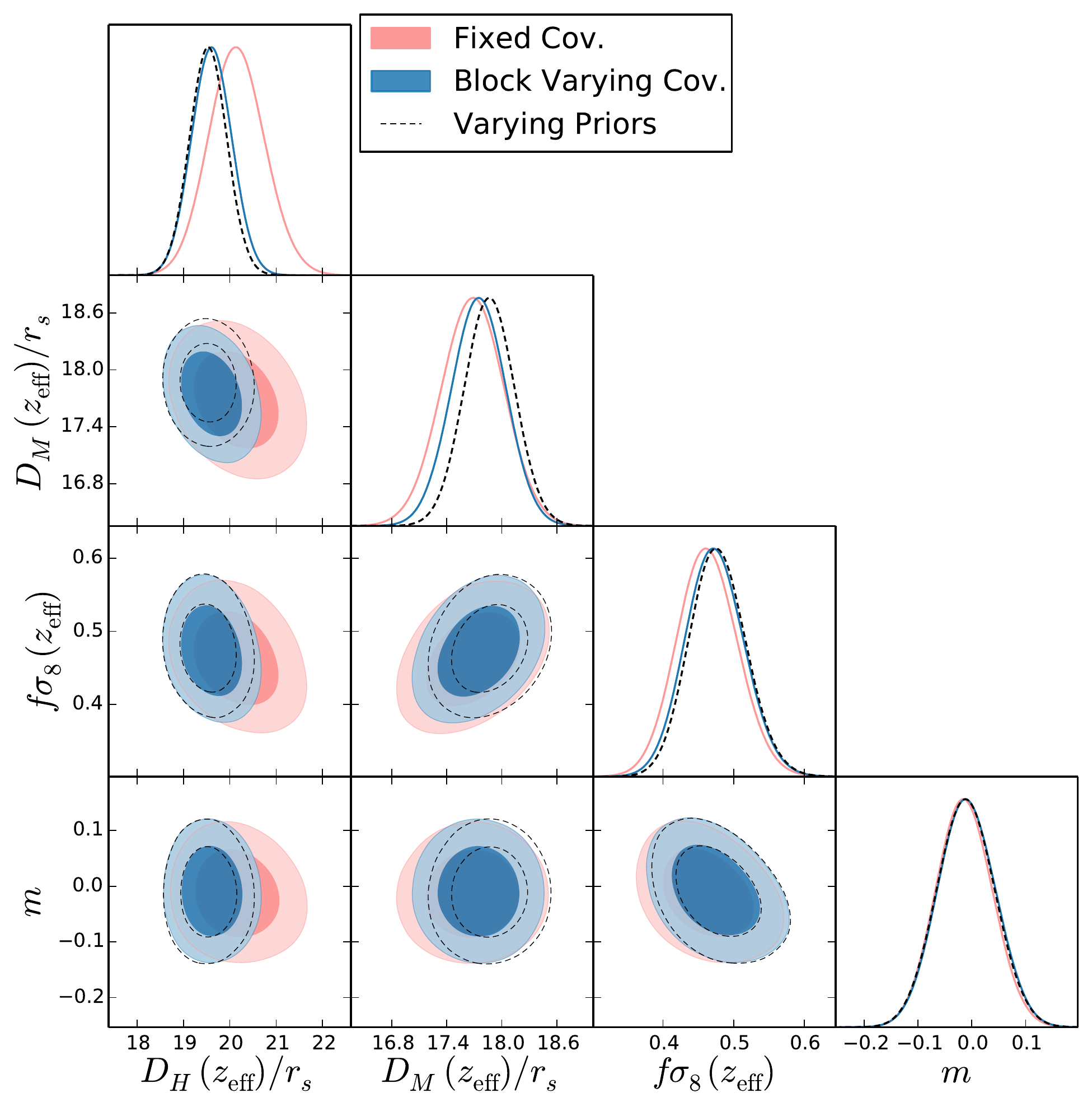}

    \caption{Posteriors drawn by analysing the eBOSS LRG data using the $PD$ approach. The different colors show two matrix architectures described in Fig.~\ref{fig:covtype}, the `fixed' covariance (pink contours), and the block- varying covariance (blue). For comparison we also display the pre-recon only analysis using varying priors (see text for details). The corresponding plot for the $DD$ approach is not shown, as it looks qualitatively very similar to the $PD$ approach.}
    \label{fig:effect_offdiag_FSalpha}
\end{figure}

Fig.~\ref{fig:effect_offdiag_FSalpha} displays the effect of the covariance architecture choice on the cosmological parameters (compressed variables) posteriors of the eBOSS LRG data, for the $PD$ case. The `fixed' covariance yields larger errors in the BAO parameters than the `block varying' case, as anticipated by Fig.~\ref{fig:FSpriors}. However, this is just by chance. If the survey had observed a different realization (or region) of the Universe with the same underlying cosmological parameters, the obtained posteriors could have been equal, broader or tighter compared to those provided by  the `block varying' choice.

We conclude that either the `block varying' or the `diagonal varying' covariances should be used for the $PD$ and $DD$ cases instead of the `fixed' covariance case. These two cases yield cosmological posteriors which are in line with the posteriors obtained from the $PP$ case. Using the `fixed' covariance architecture can lead to a significant $\sim50\%$ over or underestimation of the errors, which goes unnoticed and undiagnosed unless the complementary $PP$ analysis is performed. 

\subsection{Impact of the broadband model for the simultaneous BAO and full shape analysis}

\begin{figure}[ht]
    \centering
    \includegraphics[scale=0.5]{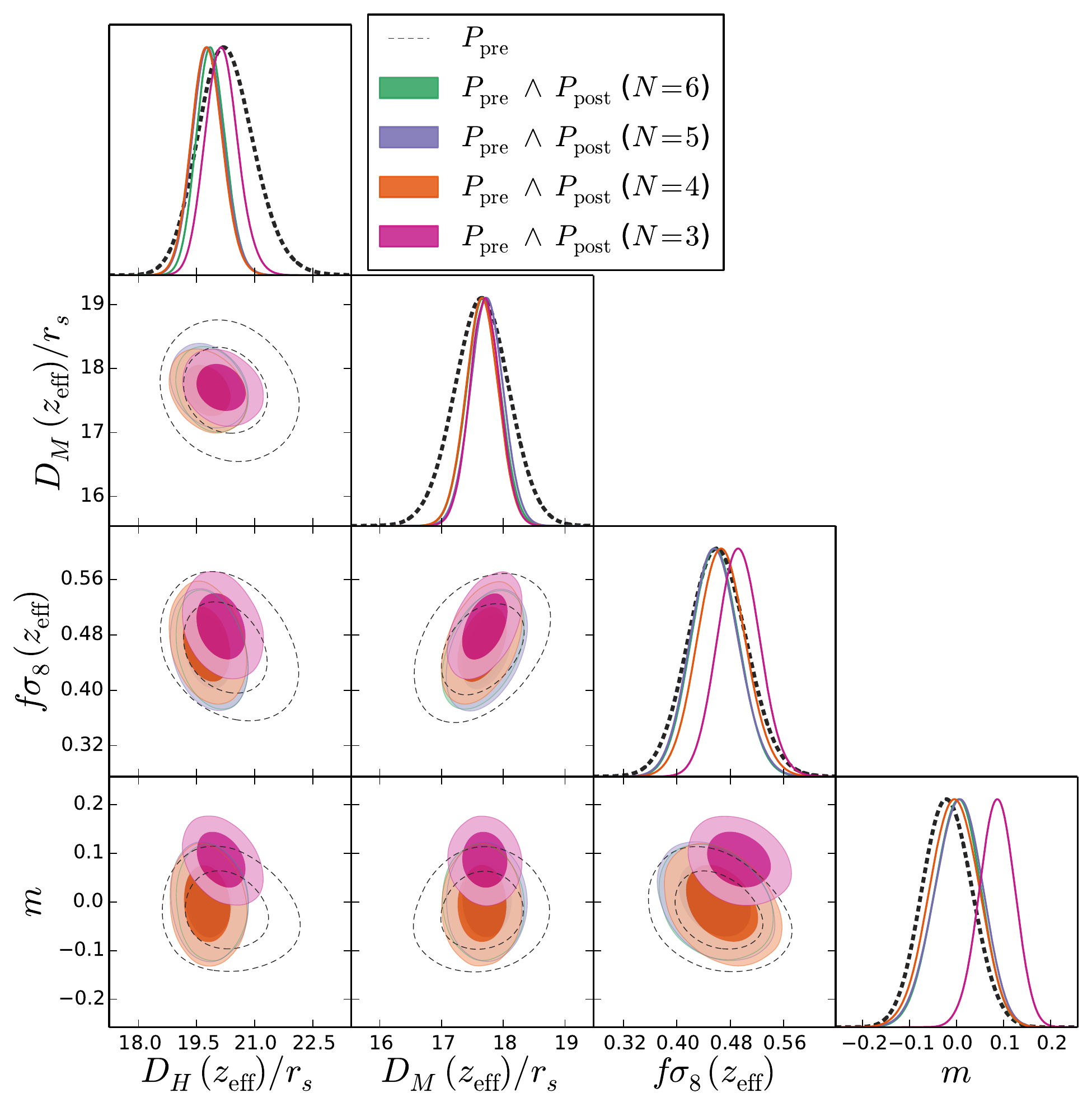}
    \caption{Posteriors drawn from the analysis of the eBOSS LRG data using the $PP$ approach. Different colors display the effect of changing the amount of polynomial terms, $N=3,\,4,\,5,\,6$ in the BAO model (see Eq. \ref{eq:poly}). For reference also the results from the pre-recon catalogue are shown for reference. }
    \label{fig:order_poly}
\end{figure}

We aim to test the impact of the order of polynomials used to describe the broadband term of the BAO when taking the $PP$ approach. This refers to the minimum $N$ value needed when considering the $N+1$ free parameters per galactic cap and per multipole in Eqs. 22-23 of \citep{Gil-Marin:2020bct},
\begin{equation}
 \label{eq:poly}   P^{(\ell)}_{\rm sm}(k)=B^{(\ell)}P^{(\rm sm)}_{\rm lin}(k)+\sum_{i=1}^N A_i^{(\ell)}k^{2-i},
\end{equation}
where $P^{(\ell)}_{\rm sm}(k)$ is the full non-linear smoothed broadband function, $P^{(\rm sm)}_{\rm lin}(k)$ is the linear smoothed function, and $\{B^{(\ell)},\, A_1^{(\ell)},\,A_2^{(\ell)}\,\ldots\,A_N^{(\ell)}\}$ are the free parameters of the models, which are different for each $\ell$-multipole and for each galactic cap. 

Already de Mattia et al. \cite{demattiaetaleboss21} stressed, for the $PP$ approach, the necessity of using a sufficiently large $N$ to avoid couplings between the BAO and full shape broadbands in the joint covariance. This minimum $N$ value could actually be different from the minimum $N$ used when considering the BAO approach alone. Indeed, the effects on the BAO peak caused by changing $N$ could be small, and still have larger effects in other broadband parameters, such as $f\sigma_8$ and $m$.

Fig.~\ref{fig:order_poly} displays the posteriors for the eBOSS LRG data on the $PP$ method for different values of $N$, from $N=3$ (pink contours) up to $N=6$ (green contours). For reference, also the pre-recon results are included in dashed black lines. We observe how the centers of the posteriors are significantly shifted and their errors under-estimated for the $N=3$ case. As we increase the order of the polynomial, for $N\geq4$ we see an excellent agreement among them and also with the pre-recon results. 

We conclude that for the specific case of the eBOSS LRG geometry and volume, considering $N\geq4$ is sufficient given the statistical errors. For the results presented in \S\ref{sec:results} we have used $N=5$.

\subsection{Covariance matrix estimation}\label{sec:cov_estimation}
So far we have not mentioned how the covariance matrix (or their cross-correlation coefficients) are estimated from the mocks. We remind the reader that we use 1000 mock realizations of the survey, and a data-vector whose size is 95 elements for the $PP$ approach. Because of the limited number of mock realizations when estimating the covariance, consisting of $95\times95$ elements, we must apply some corrections. The $PP$ approach is the one with the largest data-vector among those considered in this paper, and therefore the corrections valid for this approach will be also valid for the other two. 

In order to infer the best-fitting parameters along with their posteriors for each realization, we run a Monte Carlo Markov Chain (MCMC) based on the $\chi^2$ and likelihood functions, defined respectively as, 
\begin{eqnarray}
\label{eq:chi2}    \chi^2&=&\sum^p_{i=1} [D_i-M({\bf m})]^t C^{-1} [D_i-M({\bf m})], \\
\label{eq:L}     \mathcal{L} &\propto& \exp\left[{-\frac{1}{2}\chi^2}\right],
 \end{eqnarray}
where $D_i$ is the data-vector (for the $PP$ approach it is the power spectrum pre- and post-recon measurement for the different $k$-bins and multipoles, with $p=95$, per galactic cap); $M({\bf m})$ is the model function which depends on a series of parameters, ${\bf m}$; and $C$ is the covariance matrix. We define the covariance matrix estimator, $\hat{C}$ from the $n=1000$ realizations as,
\begin{equation}
        \hat{C}_{ij}=\frac{1}{n}\sum_{k=1}^n (d_i^{(k)}-\mu_i)(d_j^{(k)}-\mu_j),
    \label{eq:C}
\end{equation}
where $\mu_i$ is the mean value of the realizations for the element $d_i$, 
\begin{equation}
    \mu_i=\frac{1}{n}\sum_{k=1}^n d_i^{(k)}.
\end{equation}
The naive approach would be to apply Eq.~\ref{eq:C} directly into Eqs.~\ref{eq:chi2}-\ref{eq:L}. However, this procedure yields a biased estimator of the likelihood, as the inverse of Eq.~\ref{eq:C} is a biased estimator of the inverse of the covariance matrix. 

Several methods attempt to correct for this systematic error. The Hartlap-Simon-Schneider method \cite{Hartlap:2006kj} (hereafter Hartlap method) proposes to include a factor after the inversion of the covariance estimator of Eq.~\ref{eq:C} to compensate for the systematic bias. Thus, an unbiased estimator of the inverse of the covariance matrix reads as, 
\begin{equation}
    \widetilde{C}^{-1} = \frac{n-p-2}{n-1}\hat{C}^{-1}.
\end{equation}
The Hartlap unbiased estimator for the $\chi^2$ and the likelihood are,
\begin{eqnarray}
    \widetilde{\chi}_{\rm H}^2&=&\sum^p_{i=1} [D_i-M({\bf m})]^t \widetilde{C}^{-1} [D_i-M({\bf m})], \\
     \widetilde{\mathcal{L}}_{\rm H} &\propto& \exp\left[-\frac{1}{2}\widetilde{\chi}_{\rm H}^2\right].
\end{eqnarray}

Alternatively, Sellentin \& Heavens \cite{Sellentin_Heavens16} (SH hereafter) propose a correction motivated by  the need of marginalizing over the true unknown covariance matrix. In \cite{Sellentin_Heavens16} the authors do so analytically assuming that the matrix follows a Wishart distribution. The resulting sampling distribution does not follow Gaussian statistics, but a modified $t$-distribution, with wider wings and a narrower core. The SH approach takes the inverse covariance matrix from \ref{eq:C} and proposes as the unbiased likelihood estimator the following expression,\footnote{$\log$ stands for the natural logarithm.}
\begin{eqnarray}
    \hat{\chi}^2&=&\sum^p_{i=1} [D_i-M({\bf m})]^t \hat{C}^{-1} [D_i-M({\bf m})], \\
     \widetilde{\mathcal{L}}_{\rm SH} &\propto& \exp\left[-\frac{n}{2} \log\left(1-\frac{\hat{\chi}^2}{1-n}\right)\right].
\end{eqnarray}

We test these two corrections, Hartlap and SH, in the eBOSS LRG data catalogue. The resulting errors are displayed in Table~\ref{tab:cov_corr}. The table displays the results when the covariance matrix is estimated from the whole set of 1000 mock realizations, and from a reduced set of just 200 mocks, as indicated. Along with the SH and Hartlap corrections we also include the results obtained by not applying any correction (the `None' case), which corresponds to naively using the estimator of the inverse covariance, as the inverse of the covariance estimator according to Eq. \ref{eq:C}, into Eqs. \ref{eq:chi2} and \ref{eq:L}.

\begin{table}[ht]
    \centering
    \begin{tabular}{|c||c|c||c|c||c|c|}
    \hline
          & SH & SH (200) & Hartlap & Hartlap (200) & None & None (200)  \\
         \hline
         \hline
         $10^2\,\sigma_{\alpha_{\parallel}}$ & 2.030 & 2.005 & 2.029 & 2.019 & 1.923 & 1.424 \\
         $10^2\,\sigma_{\alpha_\perp}$ & 1.547 & 1.549 & 1.528 & 1.585 & 1.454 & 1.122 \\
         $10^2\,\sigma_{\widetilde{f}}$ & 6.465 & 6.356 & 6.438 & 6.442 & 6.128 & 4.636 \\
         $10^2\,\sigma_{m}$ & 5.056 & 4.824 & 5.049 & 4.885 & 4.850 & 3.849 \\
         \hline
         $\log(\mathcal{L}_{\rm max})$ & $-91.19$ & $-111.97$ & $-86.45$ & $-77.35$ & $-94.84$ & $-146.10$ \\
         \hline
    \end{tabular}
    \caption{Errors for the four cosmological related quantities, $\alpha_\parallel$, $\alpha_\perp$, $\widetilde{f}$ and $m$ derived from the $PP$ approach when different likelihood estimators are used on the eBOSS LRG data. The columns show the error values for the Sellentin \& Heavens method (SH, \cite{Sellentin_Heavens16}); the Hartlap-Simon-Schneider method (Hartlap, \cite{Hartlap:2006kj}), and the estimator resulting from not applying any correction (None). For each of these estimators we report the results when the whole set of mocks is used to estimate the covariance matrix; and when only a subset of 200 mocks is used.}
    \label{tab:cov_corr}
\end{table}

\begin{figure}[ht]
    \centering
    \includegraphics[scale=0.38]{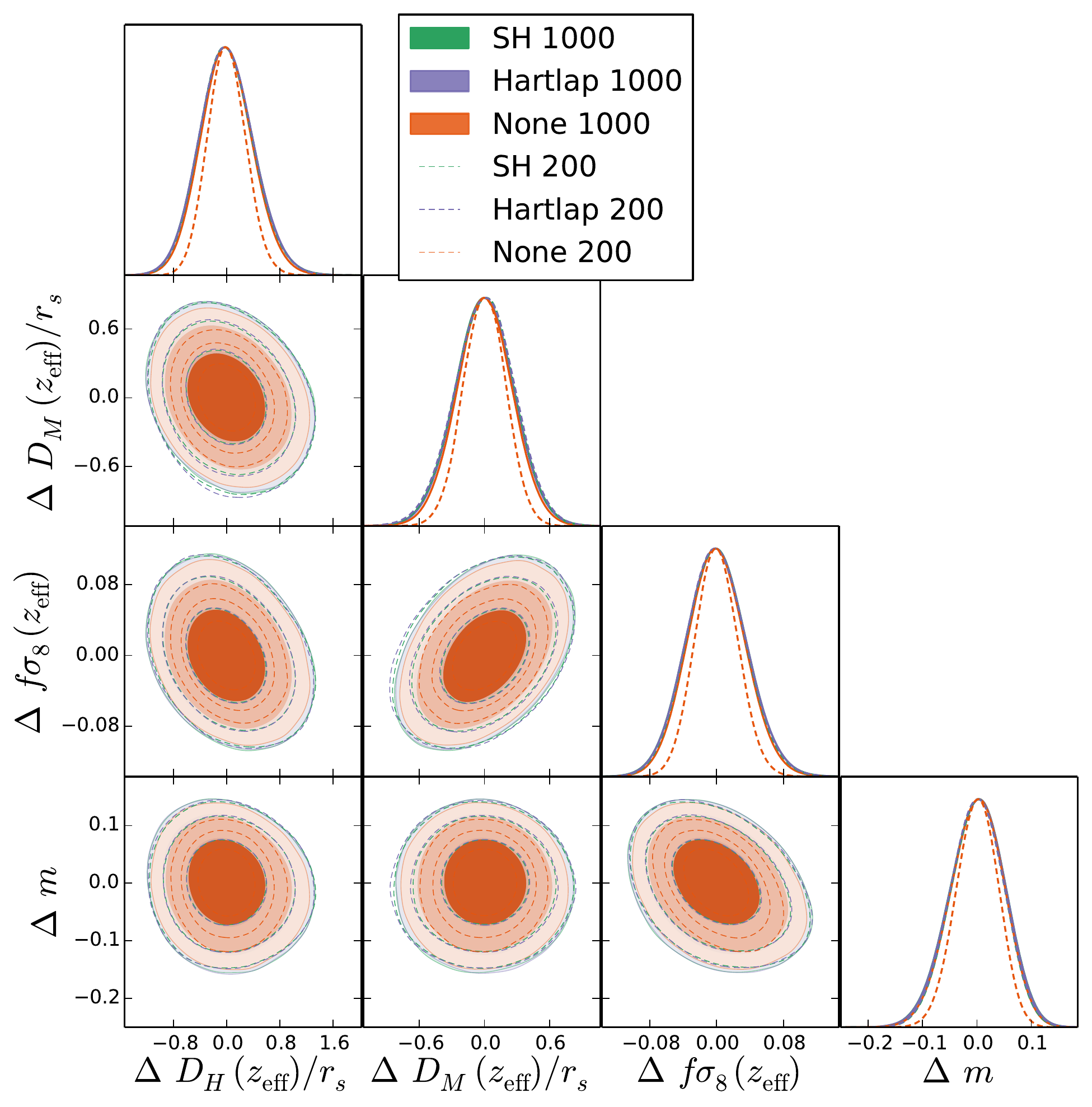}
    \includegraphics[scale=0.38]{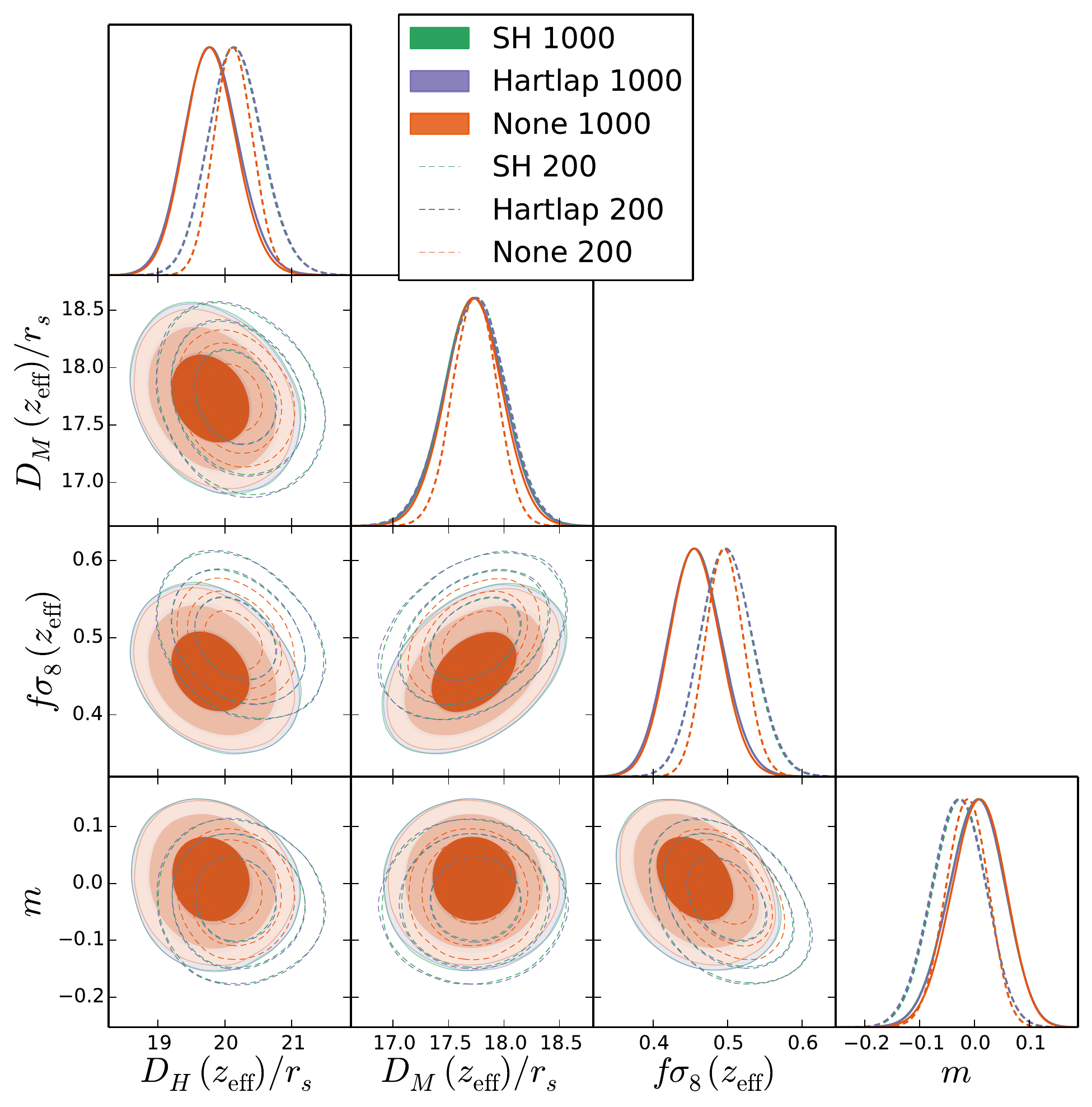}
    \caption{Impact of different estimators for the likelihood function following the $PP$ approach using three methods for estimating the posteriors of the eBOSS LRG data: Sellentin \& Heavens (SH, \cite{Sellentin_Heavens16}), Hartlap-Simon-Schneider (Hartlap, \cite{Hartlap:2006kj}), and no correction (None). In the right panel the posteriors for the cosmological derived quantities are shown; in the left panel the posteriors have been centered around the same value for a better comparison of the shape of the posteriors. The filled posteriors correspond to those analyses derived by using 1000 mocks to estimate the covariance, whereas the empty contours show the results of only using 200 mocks. The numerical values for the errors are shown in Table \ref{tab:cov_corr}.}
    \label{fig:cov_corr}
\end{figure}

The results from Table~\ref{tab:cov_corr} are visually displayed in Fig.~\ref{fig:cov_corr}. The right panel shows the actual posteriors derived for each of the analyses; whereas the left panel has centered all of them around the same position arbitrary position, for a better comparison of their shapes. 

From both Table~\ref{tab:cov_corr} and Fig.~\ref{fig:cov_corr} we see that when using the whole set of mocks, both SH and Hartlap approaches return almost identical results in terms of the errors, within $1\%$ difference, and only showing a moderate difference in the value of the maximum likelihood, $\mathcal{L}_{\rm max}$. We also report a reasonably accurate result even when no correction is applied, with a typical underestimation of the error of only $5\%$. This implies that 1000 mocks is effectively a sufficiently large number of realizations for estimating a covariance of size $95\times95$. From the right panel of Fig.~\ref{fig:cov_corr} we see that when the whole set of realizations is used to estimate the posterior (filled contours) not only the shape but the position of the posteriors is very similar. 

We compare the different likelihood estimators when only using 200 mocks to stress the differences among them in the regime where $n$ is not much larger than $p$. We report a $\sim2\%$ relative difference in the error values between Hartlap and SH. On the other hand, not applying any correction produces an error of $\sim 40\%$ with respect to Hartlap/SH. As expected, as we reduce the number of mocks when estimating the covariance, either the Hartlap or the SH correction method is needed in order to avoid significantly biasing the likelihood and underestimating the error. However, both SH and Hartlap effectively report indistinguishable results. We then conclude that for the regime of $n(=1000)>p(=95)$, and even $n(=200)>p(=95)$, they can be used indistinctly. The results from \S\ref{sec:results} were produced using SH for $PP$ and $PD$ approaches, and Hartlap for $DD$ approach, using 1000 mocks for estimating the covariance in all cases. This is because the $DD$ results were already produced from previous works, and the differences found in this section did not motivate re-running the analysis using the SH correction. 

Estimating the likelihood only using 200 mocks (instead of 1000) does not impact significantly the errors for Hartlap and SH approaches, only changing by $\leq5\%$ the diagonal errors. This is very a stable result, given the extreme low value of just 200 realizations for estimating a $95\times95$ covariance elements. However, from the right panel of Fig.~\ref{fig:cov_corr} we do observe that the center of the likelihood is shifted by $\leq \sigma/2$, equally for all the methods when using 200 instead of 1000 mocks.

\begin{figure}
    \centering
    \includegraphics[scale=0.5]{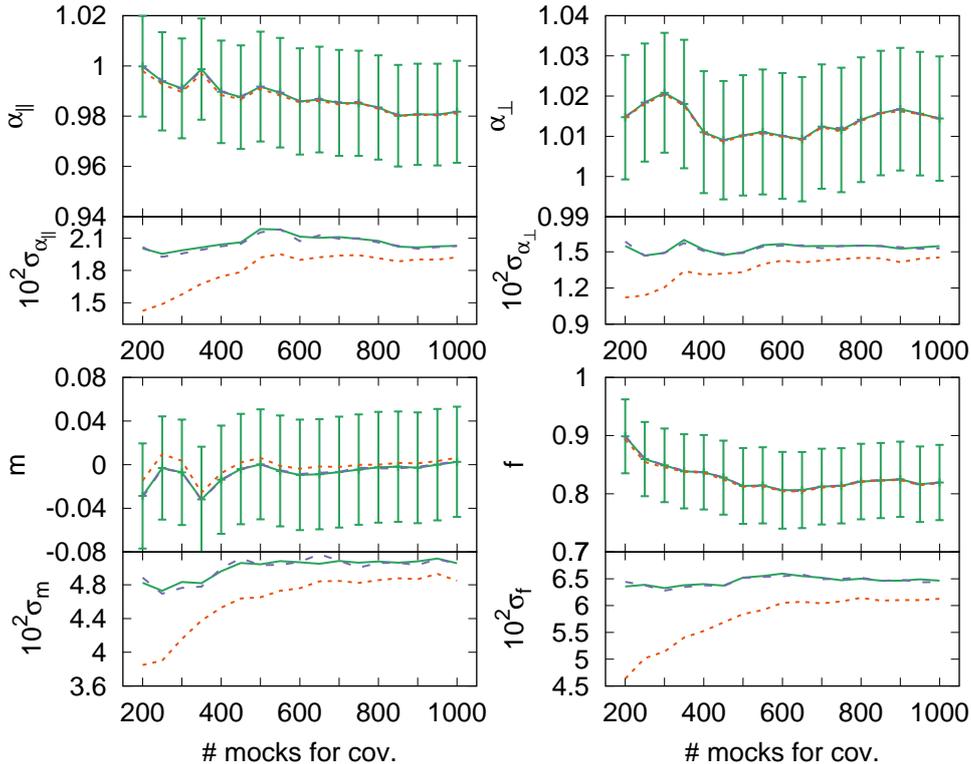}
    \caption{Impact of the number of mocks used to compute the covariance (Eq.~\ref{eq:C}) for different correction methods: Hartlap (dashed purple lines \citep{Hartlap:2006kj}), Sellentin \& Heavens (green solid lines \cite{Sellentin_Heavens16}), and no correction (orange dotted lines). The four panels show the results on the eBOSS LRG data, for the $PP$ approach (with a data-vector of 95 elements), on the cosmological parameters $\{\alpha_\parallel,\,\alpha_\perp,\,\widetilde{f},\,m\}$, for the signal (top sub-panels) and for the error (bottom sub-panels). In the signal sub-panels the error bars are only displayed for the Sellentin \& Heavens case, for clarity.}
    \label{fig:cov_evo}
\end{figure}
The effect of changing the number of mocks used to estimate the covariance is displayed by Fig.~\ref{fig:cov_evo}, for the eBOSS data using the $PP$ approach, similarly to what is shown in Fig.~\ref{fig:cov_corr}. The Hartlap (purple dashed lines) and SH (green solid) corrections are displayed, along with no correction at all (orange dotted lines). The four panels show, both the change in the signal (top sub-panel) and in their error (bottom sub-panel) as a function of the number of mocks used to estimate the covariance, for each of the four cosmological variables, as indicated. In the signal sub-panels the error bars are only displayed for the SH case for clarity. The bottom sub-panels confirm the behaviour observed in Fig.~\ref{fig:cov_corr} and Table~\ref{tab:cov_corr}, where the SH and Hartlap errors do not significantly change with the number of mocks used to estimate the covariance. However, we do observe a larger effect on the signal sub-panels, and in particular for the $\alpha_\parallel$ and $\widetilde{f}$ parameters. Both SH and Hartlap present a moderate variation of the center of the posteriors at $n=200$, which flattens to a plateau as $n$ increases, which is reached at around $n\simeq600$. The behaviour in terms of the likelihood position is surprisingly the same when no correction method is applied, as anticipated in Fig.~\ref{fig:cov_corr}. This implies that Hartlap and SH methods are accurate when accounting for the shape of the posterior, but not so much for its locus, where the posterior peaks. Consequently, even when accurate corrections to the shape of the posterior are applied, we still require to have a large number of mock realizations, of at least $\sim6$ times the entries of the data-vector $n(=600)>p(=95)$, to obtain unbiased results for the position of the posterior. 

\section{Conclusions}\label{sec:conclusions}

We have studied three different approaches for combining the full shape pre-recon and BAO-position post-recon analyses. We have explored the combination of both signals at the summary statistics level, the $PP$ approach, at the compressed variable posteriors level, the $DD$ approach, and in an hybrid manner, by combining the pre-recon summary statistics and the BAO compressed variables, the $PD$ approach (see Fig.~\ref{fig:esquema} for a visual summary). We have used the eBOSS LRG catalogues, both data and EZ-mocks, to compare the precision and the relative accuracy of these three techniques. 

We have found that only when the covariance of the compressed variables is constructed using the posteriors of the individual fits of the mocks (`varying covariance' across realizations), relevant for the $DD$ and $PD$ approaches, the results are consistent with those obtained from the $PP$ approach, which uses a `fixed covariance' across realizations. We have found no significant difference among two of the architectures studied for the $DD$ and $PD$ approaches, the `block varying' and `diagonal varying' covariances (see Fig.~\ref{fig:covtype}). On the other hand, if the covariance from the compressed variables approaches, $DD$ or $PD$, are fully derived from the ensemble of the mocks - and hence being the same for all individual realizations (the `fixed' covariance case) - the resulting error-bars can be either over- or under-estimated, depending on each particular realization.

When the appropriate varying covariance architecture is used we find very minor differences among these three approaches, reporting the $PP$ approach $\sim5-10\%$ more precision than the other two methods. This demonstrates that the $DD$ approach, widely used in the literature (see  e.g., the BOSS and eBOSS analyses \citep{alam_clustering_2017,eboss_collaboration_dr16}), is nearly optimal. When analyzing just a single realization (for e.g., only analyzing the data catalogue), the $PP$ approach is faster than the other two, as it does not require running the whole pipeline on the individual mocks. This is, the $PP$ full covariance needed for the analysis exclusively comes from the power spectra measurements, and does not require the posteriors of the individual mocks to be computed. In addition, the resulting statistics on the variables of interest for the $PP$ approach are more Gaussian than for the other two approaches (see Table~\ref{tab:moments}).

We have explored potential systematics related to the way the covariance is estimated from the mocks. We have checked that the Hartlap and Sellentin \& Heavens techniques report the same errors (only showing differences at the level of $1\%$) when estimating a $95\times95$ covariance size from 1000 mocks, as it is the case for the $PP$ approach. We also have explored the impact of modelling the broadband signal in the BAO post-recon analysis when the full shape pre-recon signal is simultaneously extracted. We have found that if the polynomial broadband model is not sufficiently high, the inaccuracies in the post-recon broadband modelling can be leaked to the pre-recon analysis, and end up systematically biasing the broadband full-shape sensitive variables: growth of structure and the shape parameters, $f\sigma_8$ and $m$, respectively. However, using moderately high polynomial, with order $\geq4$, yields no significant systematic biases for the statistical precision of the eBOSS LRG sample. 

Given the findings of this work we advocate that both $DD$ and $PP$ approaches should be used in BAO and full-shape data analyses as a cross-check mechanism, to validate the final results. However, we note that the $DD$ requires much more computational power, and hence is more time consuming, than the $PP$ approach, when we are only interested in running the pipeline on the data. The $DD$ approach needs the whole pipeline to be run in a large ensemble of mocks in order to produce the compressed-parameter covariance needed for the data analysis. This drawback makes the $PP$ approach an attractive, faster and accurate method to be used in future and on-going spectroscopic surveys.

\section*{Acknowledgements}

I would like to thank Jos\'e Luis Bernal and Licia Verde for useful comments during the development of this project.  H.G-M. acknowledges the support from `la Caixa' Foundation (ID100010434) with code LCF/BQ/PI18/11630024, the support of European Unions Horizon 2020 research and innovation programme ERC (BePreSysE, grant agreement 725327) and the IT team at ICCUB for the
help with the \textsc{Aganice} and \textsc{Hipatia} clusters, where the calculations for this project were performed. 

\appendix

\section{Cross power spectra - compressed variables terms}\label{sec:PDterms}

In this section we display the cross-correlation coefficients, $r_{ij}$ for the off-diagonal blocks of the covariance matrix for the $PD$ approach (see middle right panel of Fig.~\ref{fig:esquema}). 

\begin{figure}[ht]
    \centering
    \includegraphics[scale=0.27]{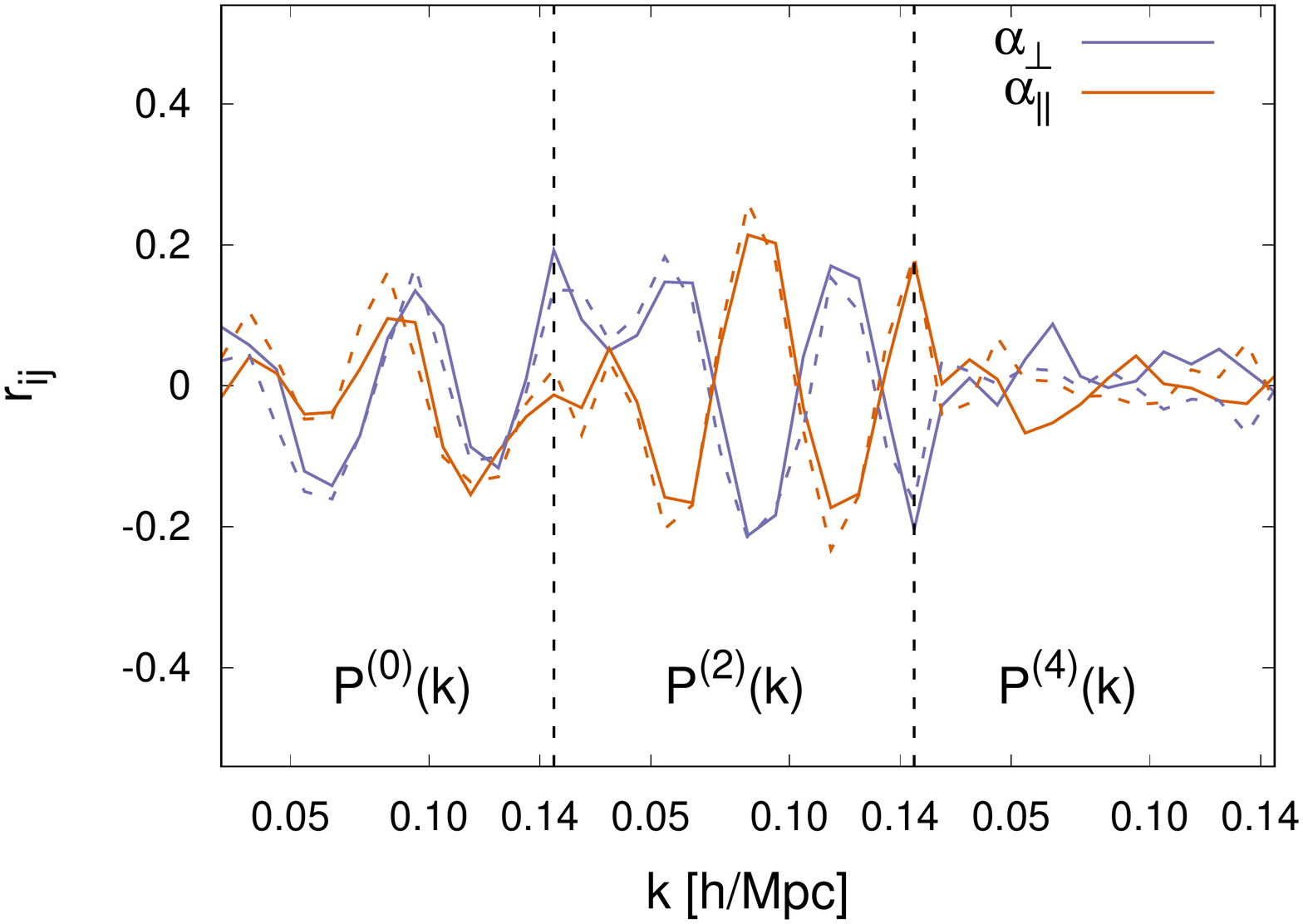}
    \includegraphics[scale=0.27]{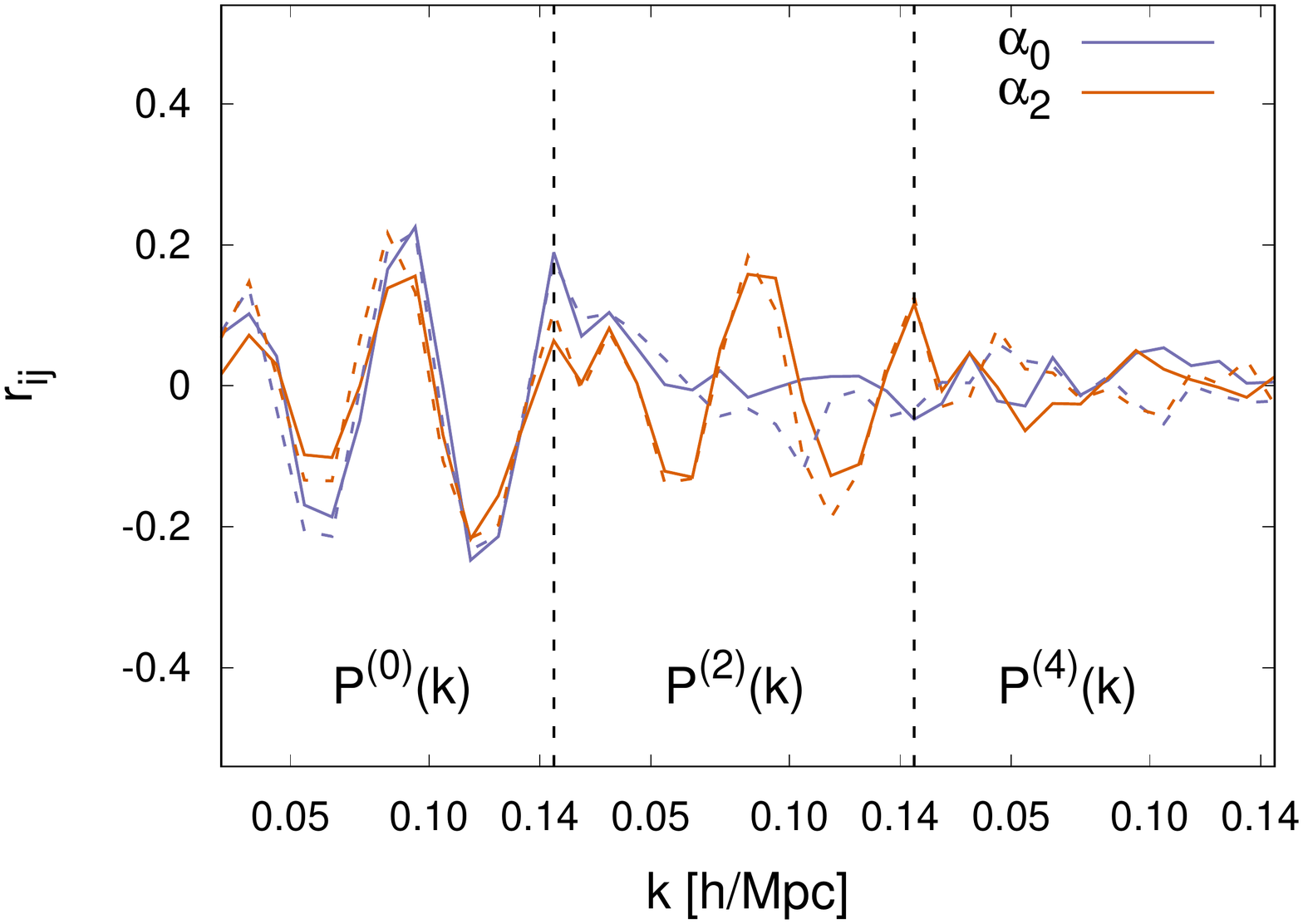}
    \caption{Cross correlation coefficients, $r_{ij}$ between the post-recon BAO variables ($\{\alpha_\parallel,\,\alpha_\perp\}$ in the left panel; $\{\alpha_0,\,\alpha_2\}$ in the right panel) and the pre-recon power spectra multipoles as a function of $k$ and $\ell$. The plots have been produced using 1000 realizations of the eBOSS LRG mocks for the northern and southern patches, corresponding to the solid and dashed lines, respectively. The solid lines in the left panel directly correspond to the off-diagonal blocks of the middle-right panel of Fig.~\ref{fig:esquema}.}
    \label{fig:Pkalphaterms}
\end{figure}

The left panel of Fig.~\ref{fig:Pkalphaterms} shows the $r_{ij}$ terms measured from the mocks ($y$-axis) between, $\alpha_\parallel$ (orange) and $\alpha_\perp$ (purple) extracted from the post-recon catalogues, and the pre-recon power spectrum $k$-bins for the three studied multipoles, represented in the $x$-axis. The solid/dashed lines represent the values for the northern/southern patches, respectively. The right panel shows also the cross-correlation coefficients, but where the $\alpha_\parallel$ and $\alpha_\perp$ variables have been rearranged into $\alpha_0=\alpha_\parallel^{1/3} \alpha_\perp^{2/3}$ and $\alpha_2=\alpha_\parallel^{3/5} \alpha_\perp^{2/5}$.

As already noted from the corresponding matrix panel in Fig.~\ref{fig:esquema}, we report an oscillatory behaviour on the $r_{ij}$ terms. In this case $r_{ij}$ represents the amount of post-recon BAO peak-position correlation with the pre-recon power spectrum at a given $k,\ell$-bin. From the left panel we see a consistent correlation between both $\alpha_{\parallel,\,\perp}$ and the $k$-bins of the monopole; whereas for the quadrupole this correlation is inverted; and for the hexadecapole is null. This behaviour can be better understood if we look at how this correlation looks in terms of the $\alpha_{0,2}$ variables in the right panel. We remind the reader that $\alpha_0$ marks the BAO position in the power spectrum monopole signal, $P^{(0)}$, whereas $\alpha_2$ marks the BAO position in the $\mu^2$-moment, defined as $P^{(\mu^2)}\equiv P^{(0)}+2/5P^{(2)}$ \citep{ross_information_2015}. In the right panel of Fig.\ref{fig:Pkalphaterms} we report a very consistent correlation between both $\alpha_0,\,\alpha_2$ and $P^{(0)}$ as expected. In fact we see that the correlation is slightly higher for $\alpha_0$ than for $\alpha_2$ because of the extra $+2/5P^{(2)}$ factor in the definition of $P^{(\mu^2)}$. On the other hand, $\alpha_0$ is very de-correlated with $P^{(2)}$, but $\alpha_2$ is highly correlated. Again, this is expected from the physical interpretation of what $\alpha_0$ and $\alpha_2$ describe. In all the cases the correlation between the post-recon $\alpha$'s and the pre-recon $P^{(\ell)}(k)$ bins oscillate between $-0.2\leq r_{ij} \leq 0.2$.
 
\begin{figure}[ht]
    \centering

        \includegraphics[scale=0.27]{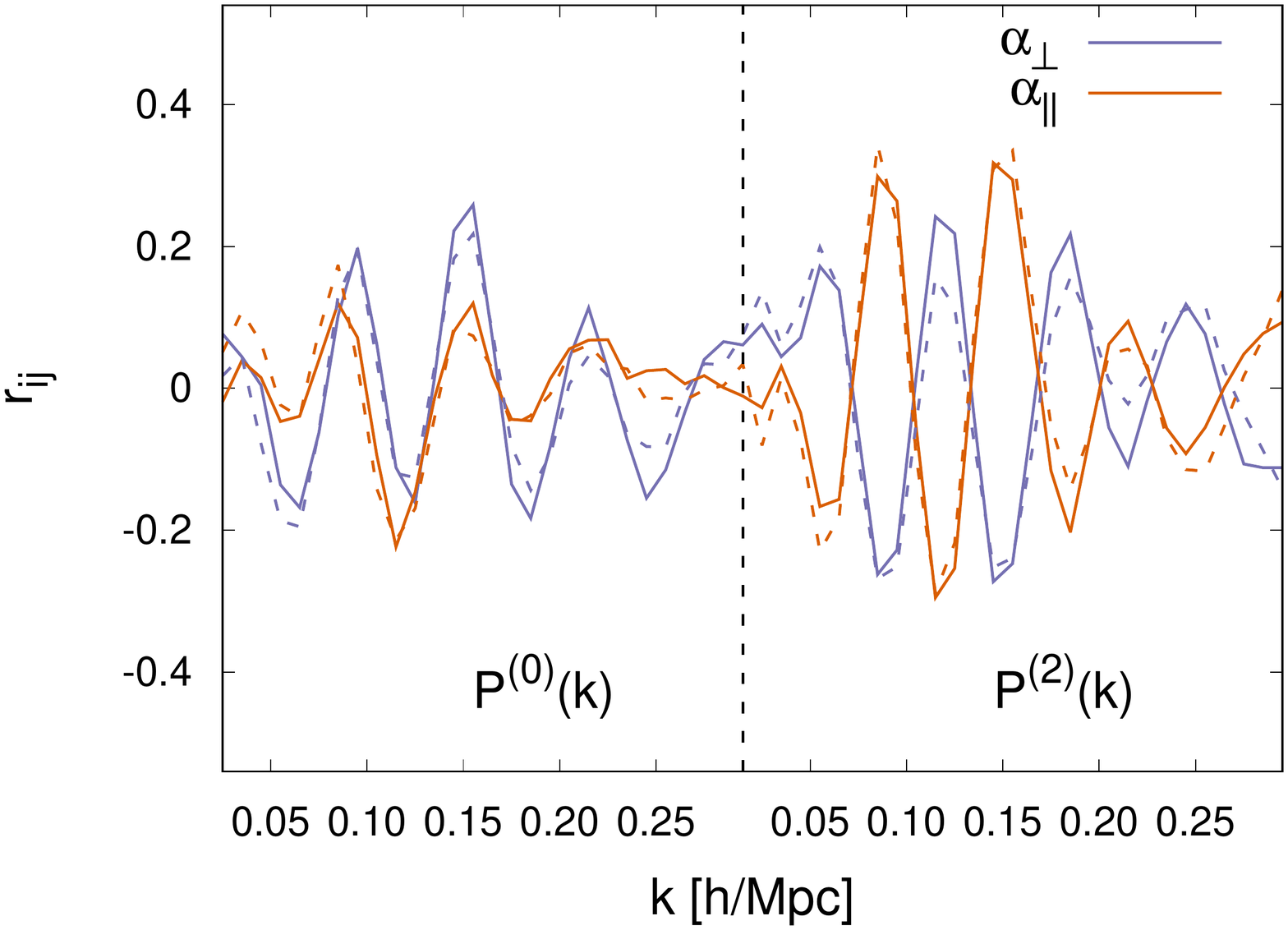}
        \includegraphics[scale=0.27]{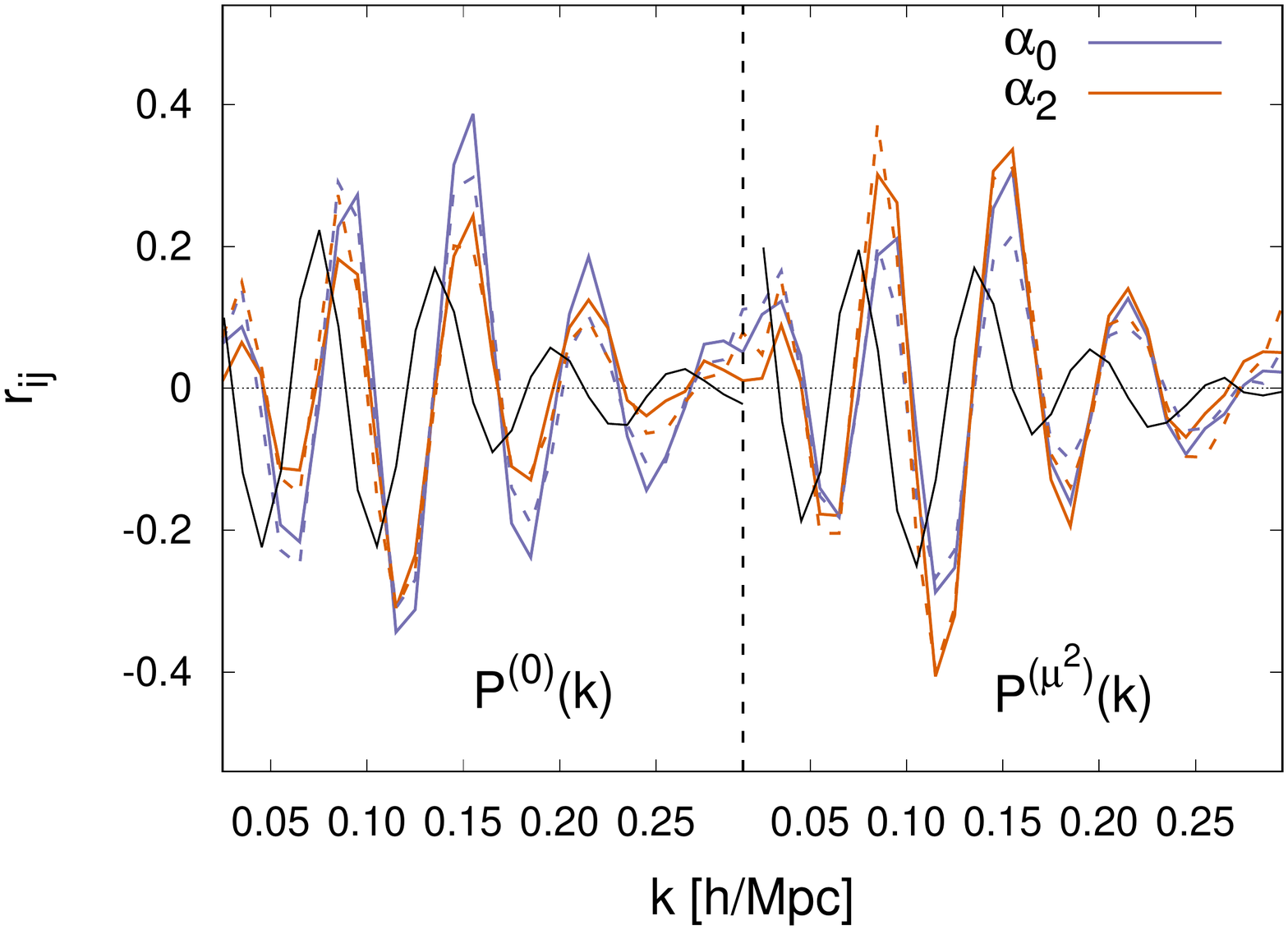}
    \caption{Same as Fig.~\ref{fig:Pkalphaterms} but cross-correlating post-recon BAO variables with post-recon power spectrum multipoles. Note that in the right panel the $x$-axis shows the correlation with respect to the monopole and the $\mu^2$-moment (see text and \citep{ross_information_2015} for definition). Additionally, for the right panel we have added in solid black line the BAO oscillatory signal pattern, linear power spectrum divided by de-wiggled linear power spectrum ($\mathcal{O}_{\rm lin}$ in arbitrary $y$-units), just to show how the peaks/wells in BAO oscillation signal correspond to $k$-bins with $r_{ij}\simeq0$. }
    \label{fig:Pkpostalphaterms}
\end{figure}

In order to better understand the oscillatory behaviour observed, we display in Fig.~\ref{fig:Pkpostalphaterms} the cross terms between post-recon BAO variables and post-recon power spectra. In this case we expect the amount of correlation to grow, as both compressed BAO variables and power spectrum multipoles are derived from the same galaxy field.  

The coefficients displayed by Fig.~\ref{fig:Pkpostalphaterms} help to understand how the BAO information is distributed along the different $k$-bins and multipoles in the power spectrum, without the effect of pre-post de-correlation. The left panel of Fig.~\ref{fig:Pkpostalphaterms} shows that the power spectrum monopole contains more information on $\alpha_\perp$ than on $\alpha_\parallel$, just the opposite behaviour than for the quadrupole. This happens because the monopole weights all $k$-directions equally, and $\alpha_\perp$ contains $N_k$ times more modes than $\alpha_\parallel$. On the other hand, the quadrupole weights the longitudinal direction with a higher weight than the transverse direction, which compensates for the unequal number of modes, and ends up being more sensitive to the BAO along the line of sight than across. For both $\alpha$'s we see a clear modulation of the correlation peaking at $k\simeq0.15\,h{\rm Mpc}^{-1}$ and demonstrating that this is the power spectrum scale more sensitive to the BAO peak-position. In the right panel the $r_{ij}$ are displayed for $\alpha_0$ and $\alpha_2$ as a function of the $k$-bins for the monopole and the $\mu^2$-moment. Again, we see a strong correlation between $\alpha_0-P^{(0)}$ and $\alpha_2-P^{(\mu^2)}$, reaching a cross-correlation of $\pm0.4$ at $k\simeq0.15\,h{\rm Mpc}^{-1}$. We also over-plot in black lines the BAO oscillatory behaviour from the BAO template, $\mathcal{O}_{\rm lin}=P_{\rm lin}/P_{\rm lin}^{(\rm sm)}$, defined as the ratio between the linear power spectrum and the de-wiggled linear power spectrum (see  eq. 20 and below from \cite{Gil-Marin:2020bct} for further details), in arbitrary $y$-axis units. We observe a $1/4$ period shift between the oscillatory behaviour of the signal template and the $r_{ij}$ terms. In fact we note the correspondence between the $r_{ij}=0$ $k$-values and the peaks of $\mathcal{O}_{\rm lin}$. In fact this is expected as the higher/lower amount of BAO information happens at those points where the derivative of $\mathcal{O}_{\rm lin}$ with respect to $\alpha_{0,2}$ is higher/lower; and the derivative with respect to $k$ happen to have the same extreme values than the derivative with respect to $\alpha_{0,2}$.

\section{Effect of the compression approach on the measured BAO variables}\label{sec:alphas}

\begin{figure}[ht]
    \centering
    \includegraphics[scale=0.27]{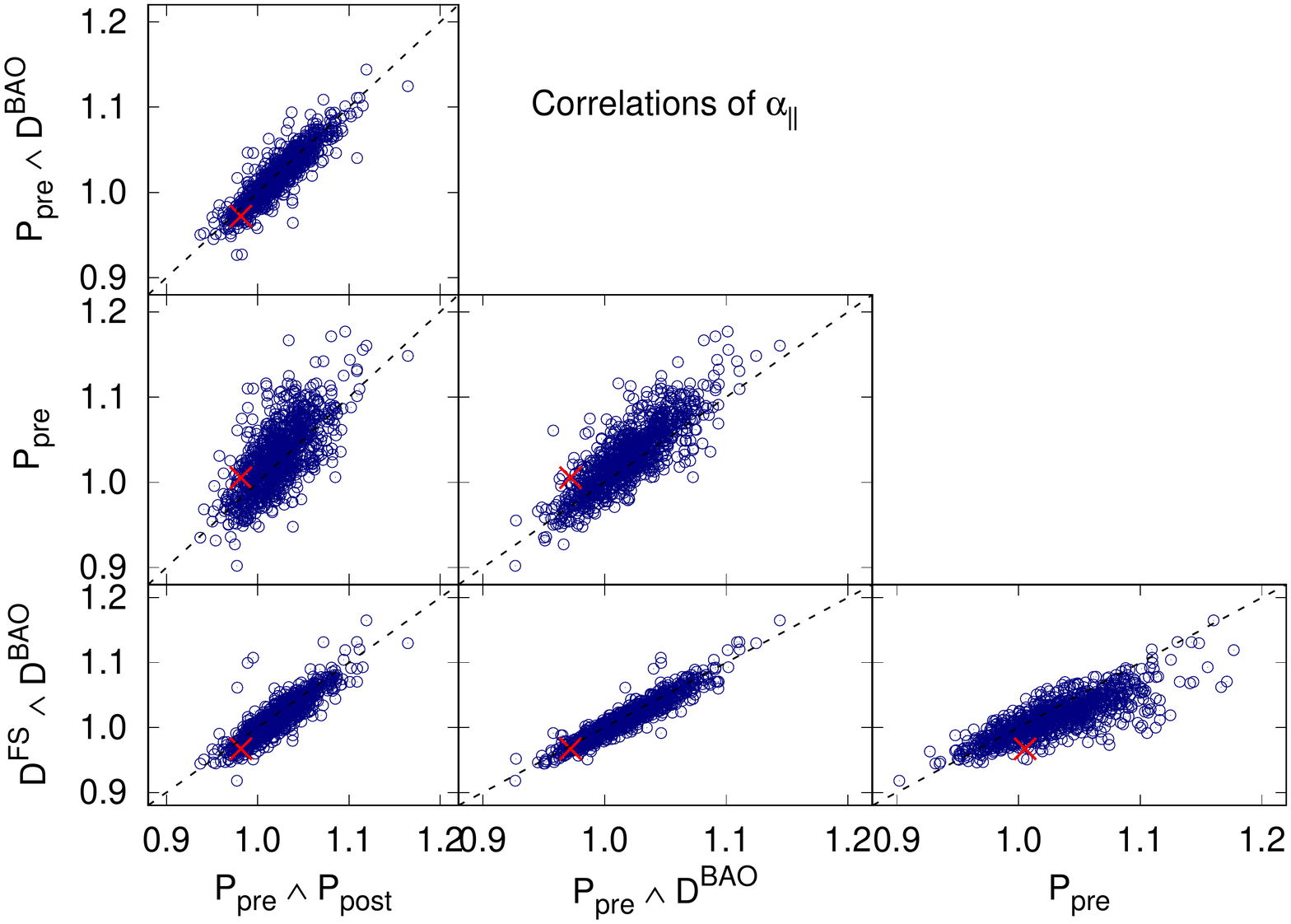}
    \includegraphics[scale=0.27]{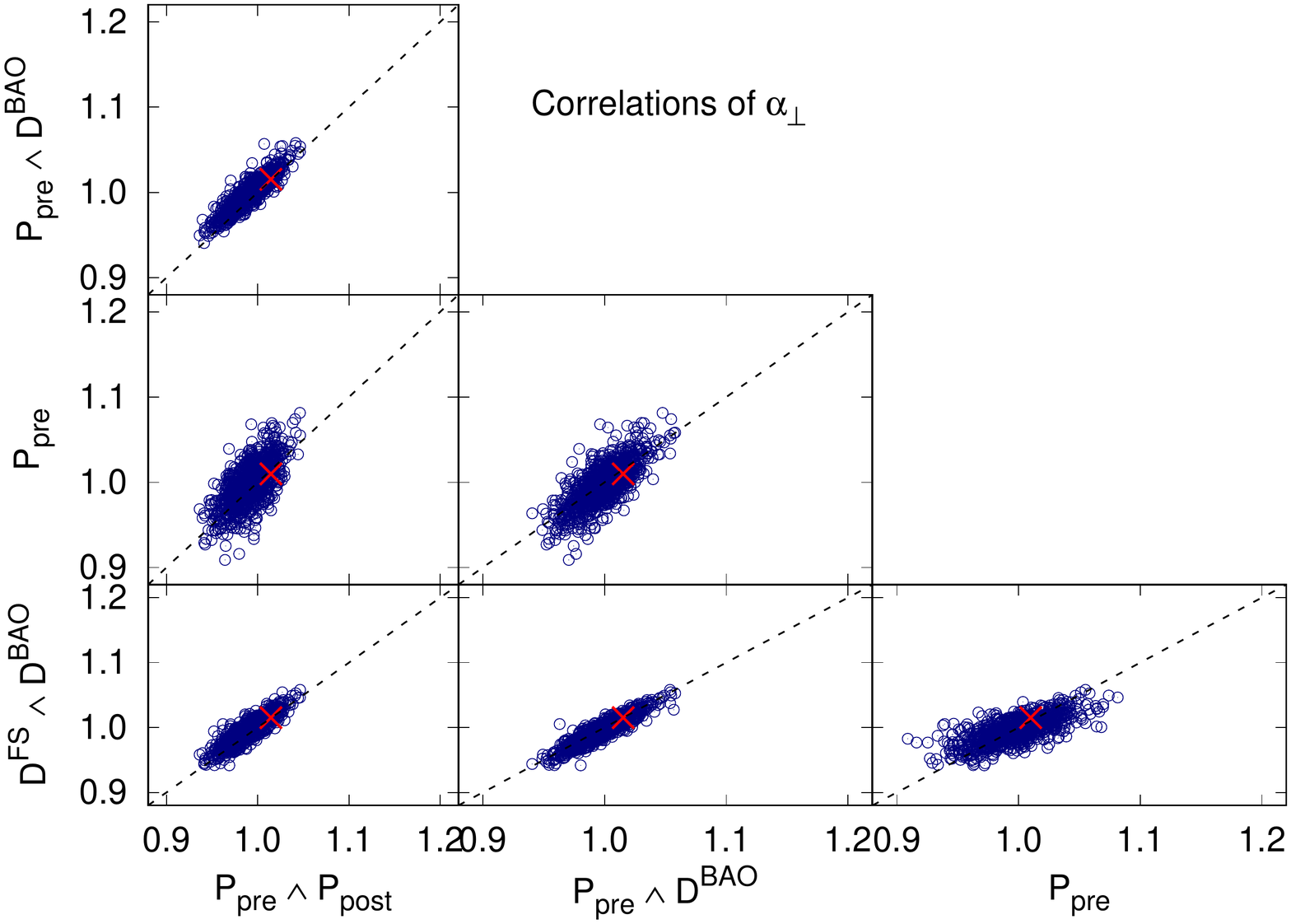}
    \caption{Best fitting values from the 1000 EZ mocks (blue dots) and data (red cross) of the eBOSS LRG sample, $0.6<z<1.0$ with $z_{\rm eff}=0.70$, for $\alpha_\parallel$ (left panel) and $\alpha_\perp$ (right panel). The different panels compare the performance for the different combining methods described in \S\ref{sec:methodology}, $PP$, $DD$ and $PD$. Also for reference the performance when only the pre-recon catalogue is used is displayed ($P_{\rm pre}$). The diagonal black dashed lines mark the equality values among methods. }
    \label{fig:signal}
\end{figure}

In this section we display the performance of the different compression methods of \S\ref{sec:methodology} on the signal of the compressed variables, $D^{\rm FS+BAO}$, focusing on the BAO variables, $\alpha_\parallel$ and $\alpha_\perp$. The left and right panels of Fig.~\ref{fig:signal} display this for $PP$, $PD$ and $DD$ compression methods, and additionally for the pre-recon only analysis ($P_{\rm pre})$ for reference, for the mocks (in blue dots) and for the data (red cross). 

The observed correlation is very high among the three compression methods. When any of these three methods is compared with the pre-recon only results we report a more scattered cloud, as expected. Still all points appear to be well aligned along the diagonal (dashed lines).

On the other hand, Table~\ref{tab:signal} displays the averaged values among all the mocks, for the different methodologies studied. Since the cosmology of the mocks is slightly different from the fiducial cosmology used to analyse the data, the $\alpha$ expected values slightly deviate from unity (also shown in the table). 

\begin{table}[ht]
    \centering
    
    \begin{tabular}{|c|c|c|}
    \hline
         & $(\alpha_\parallel-1)\times10^2$ & $(\alpha_\perp-1)\times10^2$  \\
         \hline
         \hline
        Expected & $+0.09$ & $-0.04$ \\
        \hline
        $D^{\rm FS} \wedge D^{\rm BAO}$ & $+1.55$ & $-0.37$ \\
        $P_{\rm pre} \wedge D^{\rm BAO}$ & $+1.90$ & $-0.12$ \\
        $P_{\rm pre} \wedge P_{\rm post}$ & $+2.20$ & $-0.87$ \\
        \hline
        $P_{\rm post}$ & $+0.30$ & $+0.02$ \\
       $P_{\rm pre}$ & $+2.96$ & $-0.20$ \\
        \hline
    \end{tabular}
    \caption{Averaged values among the 1000 mocks best-fitting quantities for $\alpha_\parallel$ and $\alpha_\perp$. The first row shows the expected value from the mocks (different than $\alpha_{\parallel,\,\perp}=1$ because of the different cosmology between mocks and fiducial). The rest of the rows show the different approaches for combining BAO and full shape analyses described in \S\ref{sec:methodology}, $DD$, $PD$ and $PP$. For reference also the mean values obtained when the pre- and post-recon catalogues are analyzed separately. }
    \label{tab:signal}
\end{table}

We see that all the three compression methods perform similarly in terms of recovering the expected value for the $\alpha_{\parallel,\,\perp}$. As seen in other works (see for e.g., the recent \cite{ShapeFitPT} using the perturbation theory same model for the pre-recon analysis), we report less than $1\%$ shift for $\alpha_\perp$; and $1.5-2\%$ for $\alpha_\parallel$. These systematic offsets are of the same order as those which derive from considering pre-recon analysis alone. As noted in \cite{ShapeFitPT}, these systematic shifts get significantly smaller when these parameters are interpreted within the framework of the $\Lambda$CDM model, as both $\alpha_\parallel$ and $\alpha_\perp$ become tightly related. The results of Table~\ref{tab:signal} are in line with what it was found in the left panel of Fig.~\ref{fig:histograms}, where the $PP$ approach reports a $0.75-0.50\%$ smaller value for $\alpha_\perp$ with respect to the $PD$ and $DD$ approaches. These small differences may be caused by modelling systematics of the perturbation theory model used to predict the power spectrum for the pre-reconstruction catalogue.

%
%
%


\def\jnl@style{\it}
\def\aaref@jnl#1{{\jnl@style#1}}

\def\aaref@jnl#1{{\jnl@style#1}}

\def\aj{\aaref@jnl{AJ}}                   
\def\araa{\aaref@jnl{ARA\&A}}             
\def\apj{\aaref@jnl{ApJ}}                 
\def\apjl{\aaref@jnl{ApJ}}                
\def\apjs{\aaref@jnl{ApJS}}               
\def\ao{\aaref@jnl{Appl.~Opt.}}           
\def\apss{\aaref@jnl{Ap\&SS}}             
\def\aap{\aaref@jnl{A\&A}}                
\def\aapr{\aaref@jnl{A\&A~Rev.}}          
\def\aaps{\aaref@jnl{A\&AS}}              
\def\azh{\aaref@jnl{AZh}}                 
\def\baas{\aaref@jnl{BAAS}}               
\def\jrasc{\aaref@jnl{JRASC}}             
\def\memras{\aaref@jnl{MmRAS}}            
\def\mnras{\aaref@jnl{MNRAS}}             
\def\pra{\aaref@jnl{Phys.~Rev.~A}}        
\def\prb{\aaref@jnl{Phys.~Rev.~B}}        
\def\prc{\aaref@jnl{Phys.~Rev.~C}}        
\def\prd{\aaref@jnl{Phys.~Rev.~D}}        
\def\pre{\aaref@jnl{Phys.~Rev.~E}}        
\def\prl{\aaref@jnl{Phys.~Rev.~Lett.}}    
\def\pasp{\aaref@jnl{PASP}}               
\def\pasj{\aaref@jnl{PASJ}}               
\def\qjras{\aaref@jnl{QJRAS}}             
\def\skytel{\aaref@jnl{S\&T}}             
\def\solphys{\aaref@jnl{Sol.~Phys.}}      
\def\sovast{\aaref@jnl{Soviet~Ast.}}      
\def\ssr{\aaref@jnl{Space~Sci.~Rev.}}     
\def\zap{\aaref@jnl{ZAp}}                 
\def\nat{\aaref@jnl{Nature}}              
\def\iaucirc{\aaref@jnl{IAU~Circ.}}       
\def\aplett{\aaref@jnl{Astrophys.~Lett.}} 
\def\apspr{\aaref@jnl{Astrophys.~Space~Phys.~Res.}}
\def\bain{\aaref@jnl{Bull.~Astron.~Inst.~Netherlands}} 
\def\fcp{\aaref@jnl{Fund.~Cosmic~Phys.}}  
\def\gca{\aaref@jnl{Geochim.~Cosmochim.~Acta}}   
\def\grl{\aaref@jnl{Geophys.~Res.~Lett.}} 
\def\jcp{\aaref@jnl{J.~Chem.~Phys.}}      
\def\jgr{\aaref@jnl{J.~Geophys.~Res.}}    
\def\jqsrt{\aaref@jnl{J.~Quant.~Spec.~Radiat.~Transf.}}
\def\memsai{\aaref@jnl{Mem.~Soc.~Astron.~Italiana}}
\def\nphysa{\aaref@jnl{Nucl.~Phys.~A}}   
\def\physrep{\aaref@jnl{Phys.~Rep.}}   
\def\physscr{\aaref@jnl{Phys.~Scr}}   
\def\planss{\aaref@jnl{Planet.~Space~Sci.}}   
\def\procspie{\aaref@jnl{Proc.~SPIE}}   
\def\jcap{\aaref@jnl{J. Cosmology Astropart. Phys.}}

\let\astap=\aap
\let\apjlett=\apjl
\let\apjsupp=\apjs
\let\applopt=\ao

\newcommand{\etal}{et al.\ }

\newcommand{\mpc}{\, {\rm Mpc}}
\newcommand{\kpc}{\, {\rm kpc}}
\newcommand{\hmpc}{\, h^{-1} \mpc}
\newcommand{\ihmpc}{\, h\, {\rm Mpc}^{-1}}
\newcommand{\ikms}{\, {\rm s\, km}^{-1}}
\newcommand{\kms}{\, {\rm km\, s}^{-1}}
\newcommand{\hkpc}{\, h^{-1} \kpc}
\newcommand{\lya}{Ly$\alpha$\ }
\newcommand{\lyb}{Lyman-$\beta$\ }
\newcommand{\lyaf}{Ly$\alpha$ forest}
\newcommand{\lr}{\lambda_{{\rm rest}}}
\newcommand{\bF}{\bar{F}}
\newcommand{\bS}{\bar{S}}
\newcommand{\bC}{\bar{C}}
\newcommand{\bB}{\bar{B}}
\newcommand{\vdF}{{\mathbf \delta_F}}
\newcommand{\vdS}{{\mathbf \delta_S}}
\newcommand{\vdf}{{\mathbf \delta_f}}
\newcommand{\vdn}{{\mathbf \delta_n}}
\newcommand{\vdC}{{\mathbf \delta_C}}
\newcommand{\vdX}{{\mathbf \delta_X}}
\newcommand{\xrei}{x_{rei}}
\newcommand{\lrmin}{\lambda_{{\rm rest, min}}}
\newcommand{\lrmax}{\lambda_{{\rm rest, max}}}
\newcommand{\lmin}{\lambda_{{\rm min}}}
\newcommand{\lmax}{\lambda_{{\rm max}}}
\newcommand{\hi}{\mbox{H\,{\scriptsize I}\ }}
\newcommand{\heii}{\mbox{He\,{\scriptsize II}\ }}
\newcommand{\vp}{\mathbf{p}}
\newcommand{\vq}{\mathbf{q}}
\newcommand{\vxperp}{\mathbf{x_\perp}}
\newcommand{\vkperp}{\mathbf{k_\perp}}
\newcommand{\vrperp}{\mathbf{r_\perp}}
\newcommand{\vx}{\mathbf{x}}
\newcommand{\vy}{\mathbf{y}}
\newcommand{\vk}{\mathbf{k}}
\newcommand{\vR}{\mathbf{r}}
\newcommand{\tdtwo}{\tilde{b}_{\delta^2}}
\newcommand{\tstwo}{\tilde{b}_{s^2}}
\newcommand{\tbthree}{\tilde{b}_3}
\newcommand{\tadtwo}{\tilde{a}_{\delta^2}}
\newcommand{\tastwo}{\tilde{a}_{s^2}}
\newcommand{\tabthree}{\tilde{a}_3}
\newcommand{\vnabla}{\mathbf{\nabla}}
\newcommand{\tpsi}{\tilde{\psi}}
\newcommand{\vv}{\mathbf{v}}
\newcommand{\fnl}{{f_{\rm NL}}}
\newcommand{\tfnl}{{\tilde{f}_{\rm NL}}}
\newcommand{\gnl}{g_{\rm NL}}
\newcommand{\orderfour}{\mathcal{O}\left(\delta_1^4\right)}
\newcommand{\SDSSPF}{\cite{2006ApJS..163...80M}}
\newcommand{\PF}{$P_F^{\rm 1D}(k_\parallel,z)$}
\newcommand\ionalt[2]{#1$\;${\scriptsize \uppercase\expandafter{\romannumeral #2}}}%
\newcommand{\vxone}{\mathbf{x_1}}
\newcommand{\vxtwo}{\mathbf{x_2}}
\newcommand{\vRot}{\mathbf{r_{12}}}
\newcommand{\cm}{\, {\rm cm}}

\bibliographystyle{JHEP}
\bibliography{BAO_RSD_combined}

\end{document}